\documentclass[9pt,twocolumn,twoside,final]{pnas-new}
\templatetype{pnasresearcharticle} 
\setboolean{displaywatermark}{false}

\usepackage{derivative}

\usepackage{nameref}
\usepackage[capitalise]{cleveref}
\crefname{section}{section}{sections}
\Crefname{section}{Section}{Sections}

\newcounter{subfigure}[figure]
 
\newcommand{\sublabel}[1]{\refstepcounter{subfigure}\label{#1}}

\providecommand{\noopsort}[1]{}
\usepackage{filecontents}
\usepackage{xr}
\usepackage{pdfpages}

\begin{document}

\title{Heterogeneity drives plasmid maintenance in large microbial communities}%

\author[a,b,1]{Johannes Nauta}
\author[b,2]{Kaitlin A. Schaal}
\author[c,2]{Ying-Jie Wang}
\author[b]{James P.J. Hall}
\author[c]{Shai Pilosof}
\author[a,b,e]{Manlio De Domenico}

\affil[a]{Department of Physics and Astronomy ``Galileo Galilei'', University of Padua, Padua, Italy}
\affil[b]{Istituto Nazionale di Fisica Nucleare (INFN), Sez. Padova, Padua, Italy}
\affil[c]{Department of Evolution, Ecology and Behaviour, University of Liverpool, Liverpool, United Kingdom}
\affil[d]{Department of Life Sciences, Ben-Gurion University of the Negev, Be'er Sheva, Israel}
\affil[e]{Padua Center for Network Medicine, University of Padua, Padua, Italy}

\leadauthor{Nauta}

\significancestatement{%
  Microbial communities can rapidly adapt to changes and plasmids are key drivers of such adaptation. They transfer genes that may pose threats, for example by providing antibiotic resistance, but also provide opportunities, for example by aiding bioremediation efforts. In both cases, predicting when communities can maintain a plasmid is extremely valuable. However, recent work has mostly focused on traits of a few bacteria and plasmids, while natural microbiomes are inherently large and diverse. Here, using a generic model, we show that exactly this diversity increases the probabilities that plasmids are maintained. In fact, the inherent variability can make plasmid persistence unavoidable, even without positive selection. This offers a simple and intuitive perspective on plasmid persistence in natural microbiomes.
}

\authorcontributions{%
  Author contributions:
  J.N analyzed data; J.N. and M.D.D. performed analyses; J.N., K.A.S, Y.J.W., J.P.J.H., S.P., and M.D.D. designed research, and wrote the paper.
}
\authordeclaration{
  The authors declare no competing interest.
}
\correspondingauthor{%
  \textsuperscript{1}To whom correspondence should be addressed.\\
  E-mail:~johannes.nauta\@unipd.it
}
\equalauthors{\textsuperscript{2}K.A.S. and Y.J.W. contributed equally to this work.}


\begin{abstract}
  Microbiomes are complex systems comprised of many interacting species. Species can survive harsh or changing conditions by rapid adaptation, a process accelerated by the exchange of genetic material between different species through horizontal gene transfer.
  Conjugative plasmids are ubiquitous mobile genetic elements that mediate such exchanges both within and between species.
  Therefore, predicting whether a plasmid can invade and be maintained by a microbial community is critical, for example when assessing the risks of antimicrobial resistance gene spread in commensal or environmental microbiomes.
  However, existing theory developed to assist such predictions has generally focused on the balance among plasmid costs, benefits, and infection rates, overlooking other relevant factors such as the inherent dynamics and diversity of microbiomes.
  Here, we hypothesize that plasmid persistence in the absence of positive selection can arise purely from the heterogeneity present in large and diverse microbial communities.
  We introduce a generic model that integrates population-level dynamics with plasmid conjugation.
  Using this model, we show that we can predict plasmid maintenance, and that the probability for a plasmid to be maintained depends on traits of the plasmid, most importantly the conjugation rate, and the species abundance distribution of the community.
  Then, using both empirical abundance data and extensive numerical simulations, we demonstrate that the inherent randomness of ecological interactions and conjugation rates enables plasmid persistence --- even in the absence of positive selection.
  Our findings thus suggest that natural microbial communities are likely to maintain plasmids indefinitely, offering a new perspective on the spread, maintenance, and ubiquity of plasmids.
\end{abstract}

\dates{This manuscript was compiled on \today}
\doi{\url{www.pnas.org/cgi/doi/10.1073/pnas.XXXXXXXXXX}}

\maketitle
\thispagestyle{firststyle}
\ifthenelse{%
  \boolean{shortarticle}
}{%
  \ifthenelse{\boolean{singlecolumn}}{\abscontentformatted}{\abscontent}
}{}
\firstpage[4]{3}%

\dropcap{M}icrobial systems are among the most ancient and diverse forms of life on the planet, playing foundational roles across ecological scales, from the human gut microbiome to global biogeochemical cycles~\cite{delong2001environmental,ley2006ecological,lozupone2012diversity,bardgett2008microbial,cavicchioli2019scientists}.
The environments they inhabit change continuously, hence microbial communities must be able to adapt rapidly if they are to persist~\cite{koskella2015adaptation}.
One key interaction that facilitates adaptive responses of microbial communities to changing environments is horizontal gene transfer (HGT), through which mobile genetic elements (MGEs) mediate the rapid dissemination of genetic material across individuals and taxa~\cite{thomas2005mechanisms,heuer2007horizontal,coyte2022horizontal}.
A prominent example, with serious implications for public health~\cite{martinez2008antibiotics,allen2010call,zhang2018subinhibitory,leon-sampedro2021pervasive}, is the dissemination of antibiotic resistance genes by plasmids, which enables pathogenic microbes to survive medical treatments~\cite{svara2011evolution,lopatkin2017persistence,zwanzig2021ecology,yao2022intra,gillieatt2024unravelling}.
In contrast, plasmids may also transmit desirable traits, such as genes that enable the degradation of environmental toxins~\cite{garbisu2011assessment,garbisu2017plasmidmediated}.

While plasmids are easily maintained in a microbial community when environmental conditions exert positive selection on their encoded genes, their ubiquity is additionally supported by observations that they are maintained under neutral conditions as well, despite the costs they may incur to their hosts~\cite{lili2007persistence,sanmillan2017fitness,wein2019emergence}.
While such observations put forward explanations to resolve the ``plasmid paradox''~\cite{maclean2015microbial,carroll2018plasmid}, they have been studied only in small model communities.
Another possible solution to the paradox is that the rate of plasmid transmission is, or will evolve to be, faster than the rate of extinction driven by fitness costs~\cite{brockhurst2022ecological}.
However, these and similar mechanisms have again been examined only in small model communities~\cite{hall2016source,lopatkin2017persistence,stevenson2017gene}.
In reality, natural communities are instead complex systems comprised of many distinct species, and the proposed mechanisms may break down in these systems as host variability starts to play an increasingly important role~\cite{klumper2015broad,li2020plasmids,alonso-delvalle2021variability}.
However, it is as of yet not well understood what processes or mechanisms facilitate plasmid stability in such communities.

A likely reason for this lack of understanding is that the inherent size and complexity of microbial systems makes investigating them incredibly difficult.
For example, despite advances in large-scale sequencing methods~\cite{caporaso2012ultrahighthroughput,mitchell2018ebi}, time series of microbial population dynamics with a resolution high enough to resolve plasmid abundances are, to the best of our knowledge, simply unavailable.
Still, these advances have brought to light seemingly universal patterns of variation and diversity in species abundance distributions that must, somehow, arise from the complex ecological processes that drive microbial dynamics~\cite{shoemaker2017macroecological,shade2018macroecology,grilli2020macroecological,shoemaker2024investigating,camacho-mateu2024sparse,camacho-mateu2024nonequilibrium}.
As plasmids spread within a set of distinct hosts within the community, the processes that shape the macroscopical abundance patterns must therefore influence plasmid maintenance as well~\cite{klumper2015broad}.
However, it remains unclear whether new data needs to be collected to reason about plasmid maintenance in natural communities, or whether current data may suffice to answer this question.

\begin{figure*}[t]
  \centering
  \includegraphics[width=.975\textwidth]{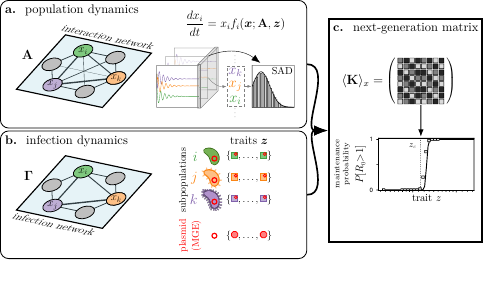}
  \caption{%
    \textbf{Predicting plasmid maintenance using the next-generation matrix.}
    \textbf{a.}~Ecological population dynamics are defined by the interaction network $\mathbf{A}$, which give --- for the plasmid-free system (see text) --- species abundance distributions (SADs).
    Note that these distributions can also be sampled from data directly.
    \textbf{b.}~Conjugation of plasmids (infection) occurs between subpopulations; this process is captured by the infection network $\mathbf{\Gamma}$.
    Infection dynamics generally depend on the traits $\boldsymbol{z}$ of the (infected) subpopulations (e.g., reduced growth rates) or the plasmids themselves (e.g., plasmid conjugation rate).
    Generally, each trait depends on a combination of the host (square) and the plasmid (red circle), but the plasmid itself may have additional traits (red squares), such as a (mean) conjugation rate.
    \textbf{c.}~By combining abundance distributions and (sampled) traits of the subpopulations, we can compute the ensemble of (random) next-generation matrices conditioned on the abundance distribution $\langle \mathbf{K} \rangle_x$, which can be used to predict plasmid maintenance.
    The plasmid maintenance probability $P[R_0>1]$ is defined as the probability that the basic reproduction number $R_0$ is larger than $1$.
    We typically find a `critical' value $z_c$ of a trait (e.g., infection rate) below which the plasmid will not be maintained (vertical dashed line).
    In addition, we find excellent overlap between analytical estimations (solid line) and numerical simulations (squares).
  }%
  \label{fig:scheme}
  \sublabel{fig:scheme:a}
  \sublabel{fig:scheme:b}
  \sublabel{fig:scheme:c}
\end{figure*}

Here, we demonstrate that species abundance distributions, of the kind readily obtained from metagenomic sequencing, suffice to predict maintenance of plasmids in microbial systems.
We introduce a generic model that captures the dynamics of interacting microbial species between which plasmids can be transferred, thereby allowing us to study the interplay between ecological interactions and epidemiological infections (i.e., plasmid conjugation).
Using this model, we then combine methods from theoretical ecology, random-matrix theory, macroecology, and epidemiology to predict when a focal plasmid is likely to be maintained (\cref{fig:scheme}).
Our results suggest that maintenance depends on the interplay between ecological interactions (which drive species abundance distributions) and plasmid conjugation (which define infection pathways).
In addition, we show that while average values of key parameters (e.g., conjugation rates) are essential, the variance of their distributions strongly influences plasmid maintenance as well.
That is to say; from the randomness that is manifested in the system parameters, a select few hosts can emerge that can, by themselves, maintain the plasmid indefinitely.
Subsequently, the plasmid may spread within the remaining members of the community under the `right conditions', such as environmental changes that favor traits conferred by hosting the plasmid.
In large systems with many distinct species, plasmid maintenance thus may depend solely on the presence of sufficient diversity, rather than on host- or plasmid-specific properties.
Our work therefore provides insight into the maintenance of plasmids within diverse microbial communities in the absence of positive selection, contributing to our understanding of plasmid ubiquity in natural microbial systems.


\section*{Results}\label{sec:results}%

\subsection{Abundance distributions predict plasmid maintenance}%
\begin{figure}[t!]
  \centering
  \includegraphics[width=\columnwidth]{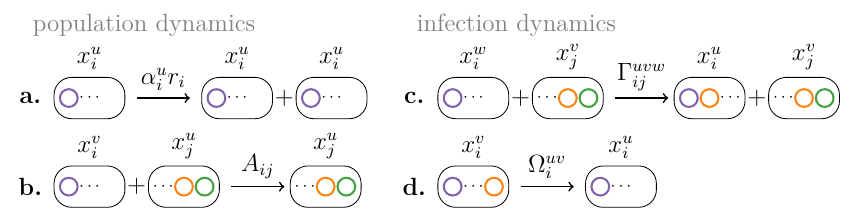}%
  \caption{%
    \textbf{Elementary reactions of a generic compartmental model with plasmids.}
    Schematic representation of the general compartmental model underlying the ecological-epidemiological dynamics model [\cref{eq:glvp}], with relevant ecological dynamics on the left, and epidemiological processes on the right.
    Colored circles represent plasmids within the host bacteria.
    Individual reactions correspond to \textbf{a.}~growth (reproduction), \textbf{b.}~interaction (competition), \textbf{c.}~infection (transmission, conjugation), and \textbf{d.}~background processes (segregation or recovery, and/or death.
    For more details on the notation and parameters, see~\nameref{sec:methods}.
  }%
  \label{fig:elementaryreactions}
\end{figure}

\label{subsec:abundancepredictsmaintenance}%
To investigate the extent to which species abundance distributions influence plasmid maintenance, we consider a generic ecological-epidemiological model that succinctly captures both population and infection dynamics~(\cref{fig:scheme}).
More specifically, we consider ecological and epidemiological dynamics to be captured by generic functions that depend on the network structures of interactions and infection.
For systems with $S$ bacterial host species and $N$ plasmids, we define a compartmental model (\cref{fig:elementaryreactions}) which defines the dynamics of abundances of each of the subpopulations $x_i^u$ as (\nameref{sec:methods}, \cref{app:sec:model})
\begin{equation}
  \label{eq:glvp}
  \odv{x_i^u}{t} = x_i^u f_i(\boldsymbol{x}; \mathbf{A}, \boldsymbol{z}) +
  \gamma_i(\boldsymbol{x}; \boldsymbol{\Gamma}, \boldsymbol{z}),
\end{equation}
where $f_i$ and $\gamma_i$ define ecological and epidemiological dynamics, respectively.
These functions depend on the (weighted) interaction networks $\mathbf{A}$ and $\boldsymbol{\Gamma}$ and the traits $\boldsymbol{z}$ of the plasmid, or host-plasmid combinations, of interest (e.g., infection rate).

\begin{figure}[ht!]
  \centering
  \begin{tikzpicture}[every node/.style={inner sep=2pt}]
    \node (PR0) {%
      \includegraphics[width=.875\columnwidth]{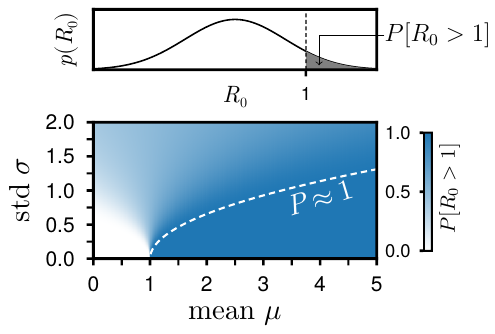}%
    };
    \node[anchor=north] (pvsbeta) at (PR0.south) {%
      \includegraphics[width=.875\columnwidth]{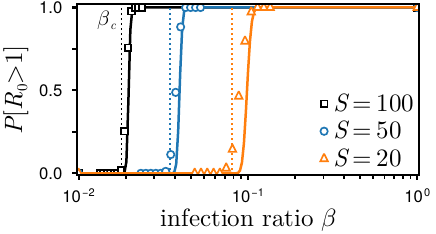}%
    };
    \node[anchor=north] (pvsa) at (pvsbeta.south) {%
      \includegraphics[width=.875\columnwidth]{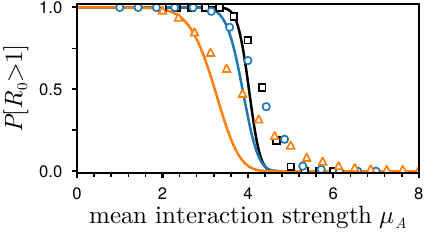}%
    };
    \node[anchor=east,font=\large] (a) at (PR0.north west) {\textbf{a.}};
    \node[anchor=east,font=\large] (b) at (a |- pvsbeta.north west) {\textbf{b.}};
    \node[anchor=east,font=\large] (c) at (a |- pvsa.north west) {\textbf{c.}};
  \end{tikzpicture}
  \caption{%
    \textbf{Predicting plasmid maintenance using abundance distributions.}
    Plasmid maintenance probability is defined as the probability that the basic reproduction number is greater than one, i.e~$R_0>1$.
    \textbf{a.}~(top)~Illustrative example of the distribution of the basic reproduction number $p(R_0)$ being a normal distribution (see text).
    The maintenance probability $P[R_0>1]$ is equal to the probability that $R_0>1$ (gray area).
    (bottom)~Maintenance probability versus the mean and standard deviation of the distribution of $R_0$.
    Dashed line indicates the line at which the probability is approximately unity --- i.e., $P[R_0>1] \approx 1$, thus below it one finds $R_0>1$ with probability $\approx 1$ and the plasmid is maintained.
    \textbf{b.}~ Effect of infection ratio $\beta$ on maintenance probability, revealing a threshold below which maintenance probability drops to (near) zero. 
    Maintenance probability versus homogeneous infection ratio $\beta = \Gamma/\Omega$ for systems with mean interaction strength $\mu_A=5$.
    Markers indicate the probability obtained by numerical integration of the dynamics, and counting the fraction of infected non-zero subpopulations (see~\nameref{sec:methods}).
    Solid lines indicate theoretical results [\cref{eq:PR0}].
    Vertical dashed lines at $\beta_c$ indicate when the maintenance probability becomes non-zero (i.e., $P[R_0>1] > \varepsilon$, with $\varepsilon = 10^{-6}$).
    For higher infection ratios, obtained by varying $\Gamma$, maintenance becomes more likely until it is guaranteed.
    \textbf{c.}~ Effect of mean interaction strength $\mu_A$ [\cref{eq:A}] on maintenance probability when infection ratio is at the critical threshold (i.e.~$\beta=\beta_c$, see~\cref{fig:mgemaintenance:a}) for systems with $S=100$ species.
    More competitive systems, obtained when increasing $\mu_A$, are accompanied by lower abundances and more functional extinction, and subsequently the maintenance probability decreases.
    Other relevant parameters are $\Omega=10^{-3}$, $\sigma_A=1.3$, $c_A = 0.2$ and $c_\Gamma=1$.
    Results are averages over 256 realizations. 
    Markers and colors in~\cref{fig:mgemaintenance:c} correspond to the same community sizes as in~\cref{fig:mgemaintenance:b}.
  }%
  \label{fig:mgemaintenance}
  \sublabel{fig:mgemaintenance:a}
  \sublabel{fig:mgemaintenance:b}
  \sublabel{fig:mgemaintenance:c}
\end{figure}
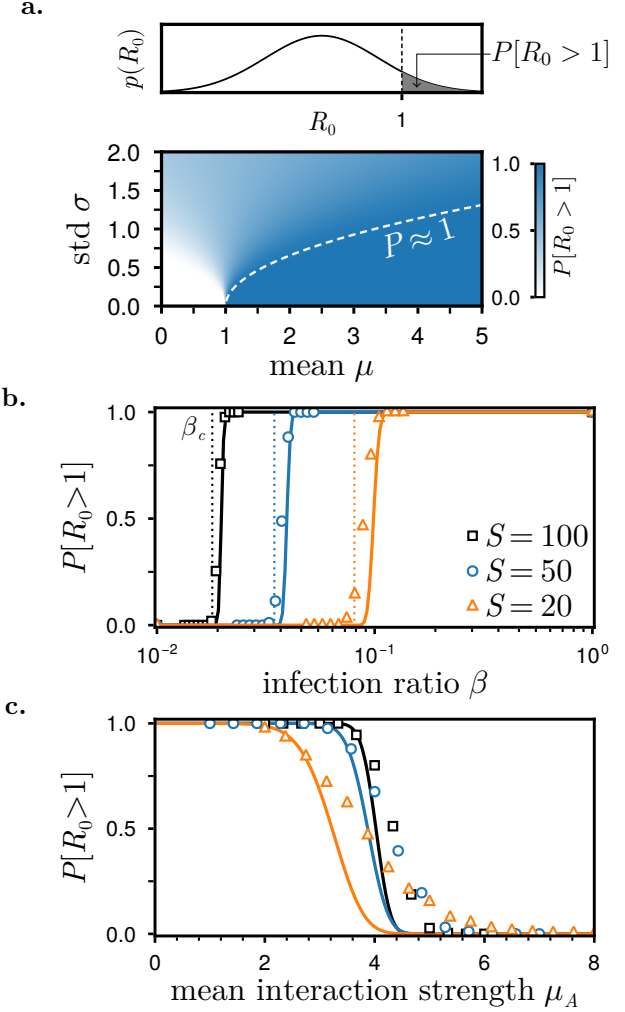

As analytical solutions to the generic dynamics of our model are generally difficult to obtain (\cref{app:sec:model}), we instead assume the interactions and infections to be random so that typical abundance distributions can be obtained.
Systems with random interaction rates are also called \emph{disordered} systems [see e.g. Ref.~\cite{mallmin2024chaotic}].
Informally, with a disordered system we mean a system for which its parameters that determine its behavior are random variables.
Note that while interactions may be chosen at random~\cite{may1972will}, it is important to realize that the population dynamics that these parameters define is deterministic.
As it turns out [see, e.g., Ref.~\cite{bunin2017ecological}], while the microscopical system parameters are random, the macroscopical distribution over species abundances is not.
In turn, we will use these abundances to construct the \emph{next-generation matrix} $\mathbf{K}$.
The eigenvalues of this matrix define the \emph{basic reproduction number} $R_0$, which indicates whether the focal plasmid is maintained~\cite{vandendriessche2002reproduction,diekmann2009construction,roberts2013characterizing,brouwer2022why} (see~\cref{fig:scheme} and \nameref{sec:methods} for more details).
Note that the species abundance distributions, as we shall show, can also be sampled directly from data, allowing us to use the framework to reason about plasmid maintenance in a more natural setting as well.

Formally, the elements of the next-generation matrix depend on species abundances $\boldsymbol{x}^\star = (x_1^\star, \ldots, x_S^\star)$, and the matrix $\boldsymbol{\beta}$ (\cref{app:sec:ngm,app:sec:R0}).
That is,%
\begin{equation}
  \label{eq:Kij}
  K_{ij} = \beta_{ij} x_i^\star,
\end{equation}
where $\beta_{ij} = \Gamma_{ij} / \Omega_i$ is the \emph{infection ratio} between the infection rate $\Gamma_{ij}$ and the recovery (plasmid segregation) rate $\Omega_i$ (\cref{fig:elementaryreactions}), and $x_i^\star$ is the abundance of species $i$ in the steady state of a system without the plasmid (see~\nameref{sec:methods}).
Specifically, the infection rate $\Gamma_{ij}$ defines the rate of infection (plasmid conjugation) between species $i$ and $j$.
Here, we fix recovery rates so that the ratio $\beta_{ij}$ acts as a direct substitute for the infection rate $\Gamma_{ij}$.
As such, we shall use infection rate and infection ratio interchangeably.
By employing the next-generation matrix formalism the problem of predicting plasmid maintenance is transformed into an eigenvalue problem.
In our specific case, the basic reproduction number $R_0$ can be computed if one has information on species abundances and infection rates of the plasmid.

\subsubsection{Abundance distributions drive plasmid persistence}%
To illustrate our proposed methodology, let us first consider a system for which species abundance distributions can be derived analytically.
First, for a focal plasmid to be maintained, we must have $R_0 \equiv \max_i \lambda_i > 1$~\cite{vandendriessche2002reproduction,brouwer2022why} (with $\lambda_i$ the eigenvalues of $\mathbf{K}$, see~\nameref{sec:methods}).
For simplicity, we assume infection to be homogeneous such that the infection ratio is the same for all species, i.e. $\beta_{ij} = \beta$.
Note that we additionally assume that the infection network, illustrated in~\cref{fig:scheme:b}, is fully connected --- that is, each species can transfer the plasmid to all other species.
Under these assumptions, the next-generation matrix $\mathbf{K}$ is a rank-one matrix and thus it has two unique eigenvalues.
The only relevant (positive) non-zero eigenvalue is in this case equal to the basic reproduction number, which reads
\begin{equation}
  \label{eq:R0simple}
  R_0 = \beta \sum_i x_i^\star
\end{equation}
From this, one can already appreciate that both high infection rates (high $\beta$) and high \emph{total} abundances increase the likelihood of the plasmid to be maintained.
What may not be immediately clear, however, is the affect that distinct abundance distributions may have on $R_0$.

To this end, we consider the case where the underlying generative model is of the disordered Lotka-Volterra type with random interactions (\cref{app:subsec:glvsads}).
In this case, species abundances follow a rectified Gaussian distribution with its moments depending on the statistics of the random interactions \cite{bunin2017ecological,galla2024generatingfunctional} [\nameref{sec:methods}, \cref{eq:rectgaussian}].
This suggests that we should consider instead an \emph{ensemble} of next-generation matrices, denoted with $\langle \mathbf{K} \rangle_x$, conditioned on the species abundance distribution from which abundances are effectively sampled (\cref{fig:scheme:c}).
Ecologically speaking, each realization of the system dynamics (i.e., a sample) will define a distinct next-generation matrix, and it is the ensemble of these matrices that we wish to analyze.
In our current example, assuming the number of species $S$ to be large, we must compute the sum of variates sampled from a rectified Gaussian.
This sum will tend to a normal distribution with mean and variance $\mu_R$ and $\sigma_R^2$ [see, e.g., Ref.~\cite{beauchamp2018numerical}].
This subsequently defines a normal distribution for the basic reproduction number as well, and hence we can compute the \emph{maintenance probability} of the focal plasmid as $P[R_0>1]$ (\cref{fig:mgemaintenance:a}), which can be extracted from the cumulative distribution function $P[R_0 > \kappa] = 1 - P[R_0 \leq \kappa]$, which reads
\begin{equation}
  \label{eq:PR0}
  P[R_0 > \kappa] = 1 - \frac{1}{2}
  \left(1 + \textrm{erf} \left[ \frac{\kappa - \mu_R}{\sqrt{2\sigma_R^2}} \right]\right),
\end{equation}
where $\mu_R$ and $\sigma_R^2$ the mean and the variance of $R_0$.
Indeed, we find excellent overlap between the closed-form solution for the maintenance probability and numerical simulations ($\kappa=1$, \cref{fig:mgemaintenance:b,fig:mgemaintenance:c}), indicating that we can predict plasmid maintenance in a model system analytically.
These results further suggest that as long as interactions are distributed such that they fall into the domain where the central limit theorem holds (i.e., their mean and variance should be finite), the basic reproduction number is (approximately) normally distributed.

However, it is well-known that abundance distributions of natural microbiomes are instead more likely to be heavy-tailed distributions~\cite{shoemaker2017macroecological,locey2016scaling,eguiluz2019scaling,wang2023origins}.
We therefore investigate the effect of strictly heavy-tailed species abundance distributions on basic reproduction numbers.
More formally, we let species abundances follow a power-law, or a Pareto distribution, with exponent $1 < \xi < 3$ (\cref{app:sec:scaling});
\begin{equation}
  p(x_i) \propto x_i^{-\xi},
\end{equation}
where we have omitted the superscript (i.e., $x_i \equiv x_i^\star$).
In such communities, $R_0$ is no longer described by a Gaussian sum and its statistics are instead controlled by the exponent $\xi$ (\cref{fig:app:R0powlaw}).

First, when $\xi > 3$ both the mean and variance exist and are finite and the classical central-limit theorem applies, for which the Gaussian approximation of \cref{eq:PR0} remains accurate.
Instead, when $2 < \xi < 3$, the mean abundance $\langle x_i \rangle$ exists and is finite, but the variance diverges.
This means that the sum $\Sigma_S = \sum_i^S x_i$ converges in distribution to a skewed $\xi$-stable distribution, replacing the aforementioned normal distribution by a $\xi$-stable one.
In this regime, the maintenance probability can still be approximated, and one finds that [\cref{eq:app:PR0stableapprox}]
\begin{equation}
  P[R_0 > 1] \propto S \beta^{\xi - 1},
\end{equation}
showing that an increase in the number of species $S$ or the infection ratio $\beta$ can outweigh a ``sub-critical'' average $R_0$ --- i.e., even when $\langle R_0 \rangle < 1$, one may still have $P[R_0 > 1] \rightarrow 1$.
In ecological terms, this means that a few extremely abundant ``super-host'' species can tip the scales towards plasmid persistence, even when the remaining bulk of the community would not sustain the plasmid on their own.

When the tail is even heavier for $1 < \xi < 2$, which may be the case in deep ocean microbiomes~\cite{eguiluz2019scaling}, the mean abundance $\langle x_i \rangle$ diverges as well and the sum $\Sigma_S$ is dominated by the single largest term $\Sigma_S \approx \max_i x_i$.
In this regime, extreme-value arguments predict that~\cite{bouchaud1990anomalous}
\begin{equation}
  \label{eq:PR0heavytail}
  P[R_0 > 1] \approx 1 - \left(1 - \beta^{\xi - 1}\right)^S
\end{equation}
which approaches $1$ extremely rapidly as either $S$ or $\beta$ grows.
This makes the average basic reproduction number conceptually meaningless, as plasmids essentially always persist because of a single extremely abundant super-host species.

Taken together, these three regimes demonstrate how the exact same plasmid can be vanishingly rare in a (from the species abundance perspective) homogeneous community, while becoming virtually inevitable in strongly heterogeneous ones.
This links community heterogeneity, mediated by the exponent $\xi$, to the fate of plasmids in microbial systems.

\subsubsection{Predicting plasmid maintenance using empirical abundance distributions}%
Whereas up to this point we have computed or sampled species abundances either from the dynamics or from a (heavy-tailed) distribution directly, our formalism also enables us to directly use abundance distributions from empirical measurements.
Here, we shall demonstrate this by using data of the gut microbiome~\cite{li2017gut}, as plasmid-mediated antibiotic resistance in such biomes presents itself as a major concern, e.g.~in clinical settings~\cite{leon-sampedro2021pervasive}.
As such natural systems like the gut microbiome are typically comprised of a diverse set of distinct species, it is important to relax the homogeneous infection assumption.
Instead, we assume a realistic distribution $p(\beta_{ij})$ over the infection ratios and let the infection network (\cref{fig:scheme:b}) be a random Erd\H{o}s-R\'enyi network with some connectivity $c_\Gamma$ (\nameref{sec:methods}).
Note that we additionally take into account the fact that the \emph{self-infection} rates $\beta_{ii}$ are typically higher than between-species infection $\beta_{ij}$~\cite{tamminen2012largescale,hu2016bacterial,alderliesten2020effect} (\nameref{sec:methods}).
For the distribution over infection rates, we assume a log-normal distribution with parameters\footnote{Note that for the log-normal distribution the mean and standard deviation are \emph{not} the same as the parameters $\mu_\beta$ and $\sigma_\beta$.} $\mu_\beta$ and $\sigma_\beta$, as recent measures of infection rates appear to align with this assumption (see~Ref.~\cite{quon2025quantifying}, \cref{app:sec:distributions}).
More formally, for the non-zero infection rates (i.e., the non-zero weights of the infection network) the distribution reads
\begin{equation}
  \label{eq:lognorm}
  p(\beta_{ij}) \sim \textrm{LogNormal}(\mu_\beta,\sigma_\beta),
\end{equation}
and, for brevity, we define the mean of this distribution as $\bar{\beta}$.

Next, using abundance data from human gut microbiomes, we substitute empirical values for $x_i^\star$ for each of the available samples, under the assumption that that these systems are in (or close to) a steady state.
It is important to realize, however, that most available datasets are compositional in nature~\cite{gloor2017microbiome}, meaning that the abundances are instead \emph{relative} abundances (i.e., $\sum_i x_i^\star = 1$).
To address the compositional nature of the data, we consider here a rescaled infection ratio $\tilde{\beta} = \bar{\beta}/B$, where $B$ is the total amount of biomass in the system (a value that is most often not available, and hence can be chosen arbitrarily).
This alleviates the necessity of knowing the absolute abundances, but gives us only information on rescaled maintenance probabilities.

\begin{figure}[t]
  \centering
  \includegraphics[width=.975\columnwidth]{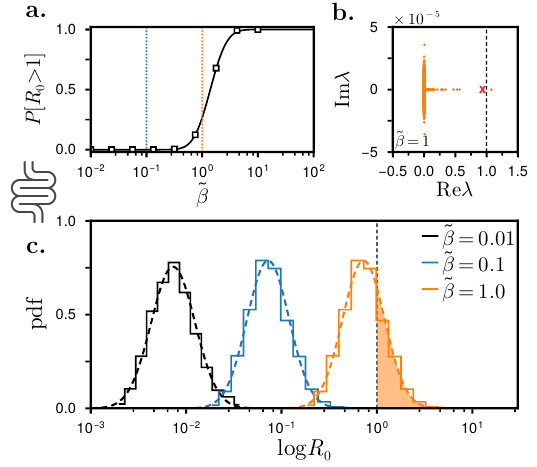}%
  \caption{%
    \textbf{Predicting plasmid maintenance using empirical data.}
    Next-generation matrices, and subsequently the basic reproduction numbers, are obtained using real data from gut microbiomes [from~Ref.~\cite{li2017gut}] and realistic infection rate distributions with $\sigma_\beta = 1$ (\nameref{sec:methods}).
    Infection networks were modeled as random Er\H{o}s-R\'enyi networks with connectivity (i.e., the probability of an edge between two nodes) $c_\Gamma = 0.5$.        
    \textbf{a.}~Plasmid maintenance probability $P[R_0>1]$ for the gut microbiome.
    The markers are obtained by computing $R_0$ from the data directly, while the line shows the theoretical approximation obtained as $p(R_0)$ is log-normal with mean $\mathbb{E}[R_0] \approx \tilde{\beta} \sum_i x_i$.
    Log-normal infection ratios have mean $\tilde{\beta} = \bar{\beta} / B$, with $B$ the total (absolute) abundance in the system.
    Note that $B$ is not measured as data contains only relative abundances (see text).
    Vertical dashed lines indicate values of $\tilde{\beta}$ for which the histogram is shown in \textbf{c} below.
    \textbf{b.}~Example plot of the complex eigenvalues $\lambda$ of a randomly chosen next-generation matrix $\mathbf{K}$ from the ensemble for $\tilde{\beta}=1$ that has $R_0 > 1$.
    Dashed line at $\textrm{Re}\lambda = 1$ indicates this threshold and the red cross indicates the expected value of $R_0$ from the ensemble.
    Note that in this example one of the eigenvalues has $\textrm{Re}\lambda > 1$ (right of dashed line), and thus the plasmid is maintained.
    \textbf{c.}~Log-normal distributions of the basic reproduction number $p(R_0)$ using empirical abundances for some values of $\tilde{\beta}$.
    Dashed lines are fitted log-normal distributions.
    Shaded area for $\tilde{\beta} = 1$ indicates $P[R_0>1]$.
    Other relevant parameters are $\sigma_\beta=1$.
  }
  \label{fig:R0gut}
\end{figure}

When $p(\beta_{ij})$ follows a log-normal distribution, we find that the basic reproduction numbers are also log-normally distributed with an expected value $\mathbb{E}[R_0] \approx \tilde{\beta} \sum_i x_i$ ($\sigma_\beta=1$, \cref{fig:R0gut}).
This result likely originates from the logarithmic scales at which (relative) abundances are distributed~\cite{grilli2020macroecological}, yet a more thorough analysis of this is considered to be out of the scope of this work.
Using the log-normal distribution, we can approximate $P[R_0 > 1]$ noting that for a log-normal distribution the cumulative distribution function is the essentially the same as in~\cref{eq:PR0} but with the substitution $\kappa \rightarrow \log \kappa$.
Our approximation of $R_0$ matches those computed from next-generation matrices using the data explicitly very well, indicating that our predictive framework can be readily used with empirically obtained abundance data (\cref{fig:R0gut}).

For gut microbiomes specifically, additional knowledge on absolute abundances $B$ is required, yet these can be inserted into our framework naturally.
By noting that abundance distributions tend to follow log-normal (or similar) distributions, one may again appreciate the significant effect of extremely abundant species on plasmid maintenance in natural microbiomes.
However, when sampling species abundances directly, it remains unclear how ecological dynamics may affect plasmid maintenance.

\subsubsection{Competition negatively affects plasmid maintenance}%
As one may expect, the above results show that the most important system variables are those that influence the distribution of basic reproduction numbers [\crefrange{eq:R0simple}{eq:PR0heavytail}].
As mentioned earlier, these are the abundance distributions themselves, but they also implicitly include the distributions over interaction and infection coefficients.
For species abundances, the above results highlight that heavy-tailed abundances may lead to the inevitability of plasmids in sufficiently diverse communities.
When abundance distributions do not have a heavy-tail, however, results are more nuanced.
For example, in the context of the generalized Lotka-Volterra model [\cref{eq:glvp}], higher average interaction strengths, which define more competitive interactions, generally reduce the likelihood of plasmid maintenance (\cref{fig:mgemaintenance:c}).
The underlying mechanism is twofold.
First, the increase in interaction strength corresponds to higher levels of competition which decrease the total abundance of the model community.
This subsequently lowers the average basic reproduction number $R_0$, thus leading to the loss of plasmids [\cref{eq:PR0}, \cref{app:sec:R0,app:sec:sads}].
Second, competition may result in \emph{functional extinction}, where the abundance of a particular species becomes so low that it effectively does not play any role in the dynamics~\cite{zeeman1995extinction,parker2009extinctiona,arnoulxdepirey2024manyspecies}.
Intuitively, species death ``disrupts'' the infection network (as host species, i.e. the nodes of the network, vanish), which diminishes the spread of the plasmid thereby leading to the plasmid being lost from the community.

\begin{figure*}[t]
  \centering
  \begin{tikzpicture}[every node/.style={inner sep=0pt,font=\Large}]
    \node[anchor=west] (dist) {%
      \includegraphics[width=.825\linewidth]{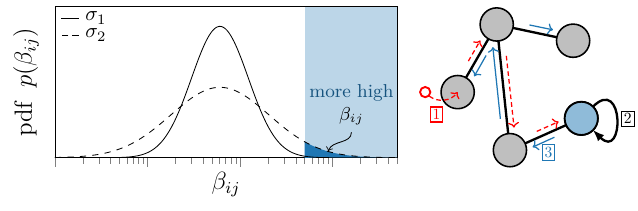}%
    };
    \node[anchor=north] (PR0) at (dist.south) {%
      \includegraphics[width=.925\linewidth]{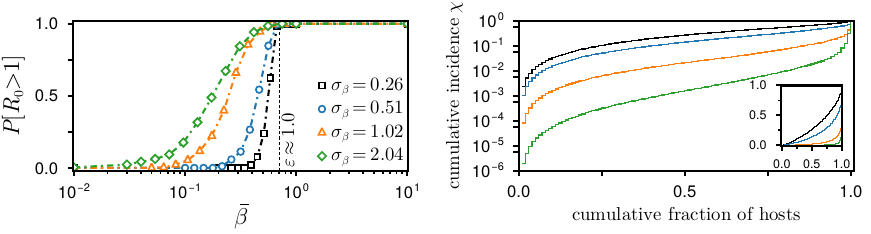}%
    };
    \node[anchor=west] (a) at (dist.north west) {\textbf{a.}};
    \node[anchor=west,xshift=2.5cm] (b) at (dist.north) {\textbf{b.}};
    \node[anchor=west] (c) at (PR0.north west) {\textbf{c.}};
    \node[anchor=center] (d) at (PR0.north) {\textbf{d.}};
  \end{tikzpicture}
  \caption{%
    \textbf{Rare hosts induce source-sink dynamics.}
    \textbf{a.}~An illustrative example of how the frequency of high infection ratios $\beta_{ij}$ depends strongly on the variance of the log-normal distribution, even when the mean $\bar{\beta}$ is the same.
    We let $\sigma_2 > \sigma_1$, so that the dark shaded area between the two curves illustrates the increased likelihood of sampling (more) high values of $\beta_{ij}$ for distribution with increased variance.
    As a result of sampling more high $\beta_{ij}$, the likelihood of sampling favorable hosts for the plasmid increases such that infecting those hosts alone is enough to be maintained indefinitely (see text).
    \textbf{b.}~Illustration of source-sink dynamics in a connected infection network $\boldsymbol{\Gamma}$, here with $5$ nodes (hosts).
    (1)~The plasmid (red circle) is introduced in a random host and initially spreads through the system (red dashed arrows) until it reaches a favorable host (blue).
    (2)~The favorable host is abundant and has, for example, a high self-infection rate sampled from $p(\beta_{ij})$ (see \textbf{a}), allowing it to self-maintain the plasmid indefinitely.
    (3)~The favorable host now acts as the source of infection for the system (blue arrows), allowing the plasmid to spread through the community.
    \textbf{c.}~Plasmid maintenance probability versus the mean infection ratio $\bar{\beta}$ for distinct log-normal distributions with increasing $\sigma_\beta$ for a system with $S=100$ species.
    As the variance increases as $\sigma_\beta$ increases, so do probabilities of plasmid maintenance as extreme traits (high infection ratios) sampled from the tails of the log-normal become more frequent enabling source-sink dynamics (as illustrated in \textbf{a.} and \textbf{b.}).
    Vertical dashed line indicates $\bar{\beta}$ beyond which $P[R_0 > 1] \approx 1$.
    \textbf{d.}~Assessing the contribution of each host on system-wide plasmid spread.
    We plot the cumulative infectious incidence $\chi$ (\cref{eq:chi}, and see~\nameref{sec:methods}) versus the cumulative fraction of hosts, identifying a select few (favorable) hosts which are responsible for most of the infection in the system.
    Hosts are ranked by increasing contribution to plasmid conjugation.
    As the variance of infection ratios increases, system-wide infection is increasingly driven by a smaller fraction of favorable (source) hosts.
    Inset shows the same figure but in linear scale.
    Parameters and colors are as in~\cref{fig:mgemaintenance} and \cref{fig:mgemaintenance:c}.
    Other relevant parameters are $d=0.1$ and $\bar{\alpha}=0.9$, $\bar{\beta} > \bar{\beta}_c$.
    Results are averaged over 256 realizations.
  }
  \label{fig:infectionsource}
  \sublabel{fig:infectionsource:a}
  \sublabel{fig:infectionsource:b}
  \sublabel{fig:infectionsource:c}
  \sublabel{fig:infectionsource:d}
\end{figure*}
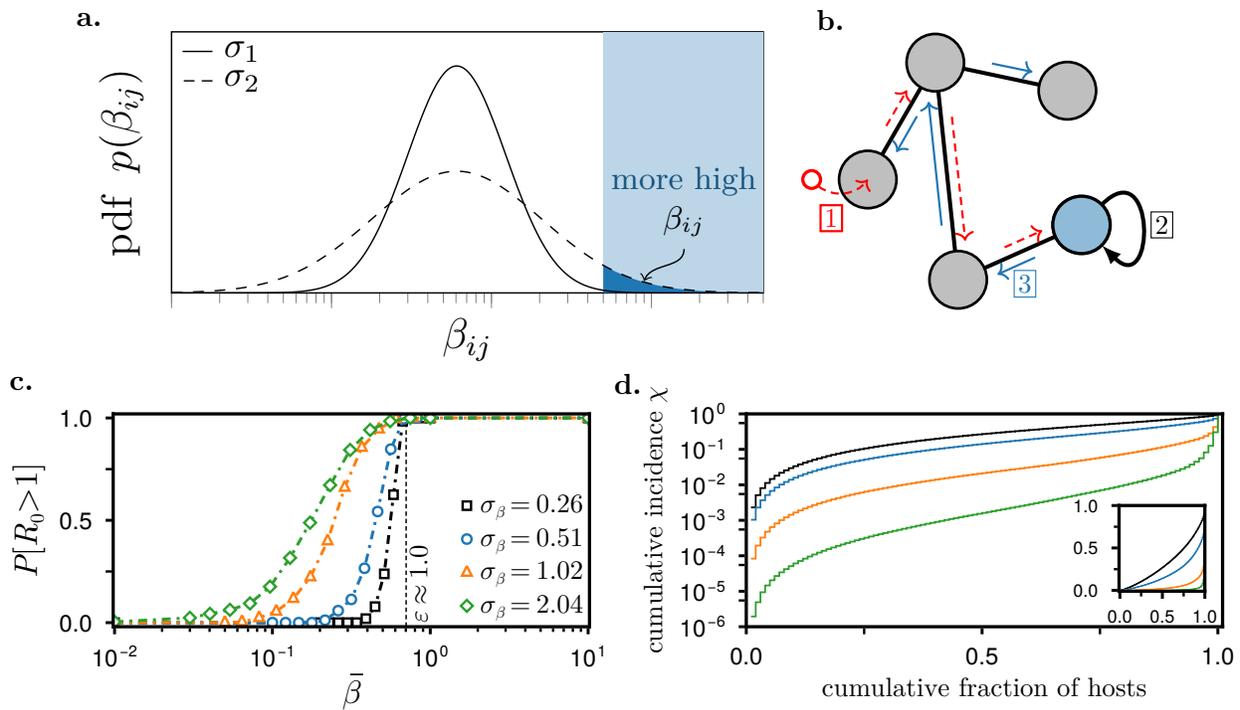

\subsection{Specific hosts as sources of infection}%
The above results suggest that, in contrast to the epidemiological mechanisms that underlie persistence of a focal plasmid (such as increased infection rates), the ecological mechanisms at play, such as competition and extinction, tend to \emph{reduce} maintenance probabilities.
In fact, most mechanisms that we considered (either ecological or epidemiological), such as explicit infection networks (see below) or reductions in growth rates when hosting the plasmid (\cref{app:sec:dilution}), typically introduce yet another avenue through which the focal plasmid can be lost.
Consequently, one may ask whether heavy-tailed abundance distributions are a necessary component for plasmids to be maintained, or whether sufficient variation in other system parameters may do the trick.

Empirical observations support the latter view, showing that even in the absence of heavy-tailed abundance distributions not all hosts may contribute equally to plasmid maintenance~\cite{degelder2007stability,kottara2018variable,li2020plasmids,alonso-delvalle2021variability}.
Moreover, similar to the extremely abundant species mentioned above, a small subset of hosts often appears to be responsible for the majority of plasmid infections in the system~\cite{hall2016source}.
This suggests that perhaps sufficient variability in conjugation rates may underlie plasmid maintenance.

To test this, we consider systems with normal-like abundance distributions [specifically the Lotka-Volterra model of~\cref{eq:glvp}], incorperating variability via the parameters of the conjugation rate distribution $p(\beta_{ij})$.
In addition, we add a random (negative) fitness effect of hosting the plasmid for a particular host, such that for some host $i$ its infected subpopulation, denoted with $y_i$, follows the Lotka-Volterra model but with growth rate $\alpha_i r_i$~\cite{bergstrom2000natural}.
Note that in order for the fitness effect to alter the system's behavior, we also need to consider \emph{diluted} Lotka-Volterra systems with a constant dilution (or death) rate $d$ (\nameref{sec:methods}), as in the undiluted case the fitness cost $\alpha_i$ has no effect (\cref{app:sec:dilution}).
For both the infection rate and the fitness effect, we align ourselves with empirical observations~\cite{alonso-delvalle2021variability,quon2025quantifying} and restrict ourselves to distributions with finite moments.
We again denote the mean, for example of the infection rate, as $\bar{\beta}$.
We explicitly consider the infection rates to be the weights of an infection network with degree distribution $p(k)$ (as in~\cref{fig:scheme:b}) with average degree $\bar{k}$.

When assuming a log-normal distribution for both infection rates and fitness effects (\nameref{sec:methods}), we can modulate the variance of the infection rates by changing the parameter $\sigma_\beta$.
When $\sigma_\beta$ increases, note that the likelihood of sampling a host $i$ that has a high infection rate $\beta_{ij}$ to some of its neighbors (or to itself) increases~(\cref{fig:infectionsource:a}).
As such, under the assumption that the initial infection network is connected (i.e., that there exists a path between any two nodes $i$ and $j$), the plasmid can ``travel'' the network until it encounters the most favorable host (or a favorable set of hosts).
Once those are infected, this ``super-host(s)'' may act as the source for any further infection (\cref{fig:infectionsource:b}).
That this is indeed the case become tangible when looking at the effect of $\sigma_\beta$ on plasmid maintenance probabilities (\cref{fig:infectionsource:c}).
Indeed, when the likelihood of sampling a super-host increases (increased $\sigma_\beta$), the maintenance probability for a given mean infection ratio $\bar{\beta}$ increases as well.
The effect is similar to those described above for heavy-tailed abundance distributions, but now depends on sufficient variability in infection rates.
That is to say, even for low \emph{average} infection rates $\bar{\beta}$, plasmids can potentially be maintained with a non-zero probability.
Thus, the randomness that is manifested in a combination of interactions, infections, and/or host fitness costs, may also underlie plasmid maintenance.
We would like to mention that this result aligns with empirical findings that indicate that host-plasmid traits related to plasmid cost and conjugation may vary substantially across taxa~\cite{degelder2007stability,kottara2018variable,li2020plasmids,alonso-delvalle2021variability}, thus increasing the odds of finding a favorable host in natural systems in which the number of species is typically large.
Note that this effect emerges even when plasmids confer a negative fitness cost on their host ($\alpha_i < 1$, \cref{app:sec:dilution,fig:app:singlehostpathogenR0:a}).

Our results further solidify the observation that only very few hosts effectively spread the plasmid --- even when maintenance is guaranteed ($\bar{\beta} > \bar{\beta}_c$, \cref{fig:infectionsource:c,fig:infectionsource:d}).
Similar to the case when abundances are heavy-tailed, system-wide maintenance appears to originate only from a handful of super-host species.
We investigate the sources of infection in our system by quantifying the total rate at which each community member infects others with the plasmid.
To this end, we define the \emph{infectious incidence} $\chi_i$ as
\begin{equation}
  \label{eq:chi}
  \chi_i = \sum_j \beta_{ij} y_j
\end{equation}
That is, the infected subpopulation of a particular host $y_i$, typically grows with $x_i\chi_i$.
After ranking the incidences of the entire population from low to high, we can obtain the cumulative incidence $\chi_\ell$ at rank $\ell$ (see~\nameref{sec:methods}), that is, how much of the system-wide plasmid transfer is accounted for by species $\ell$.
Note that such a definition for $\chi_i$ (or $\chi_\ell$) effectively measures the inequality of the total outgoing infections, in which one may recognize the similarity with other inequality measures such as the Gini coefficient in economics (see also~\cref{fig:infectionsource:d}, inset).
In doing so, we see that for systems with high variance (high $\sigma_\beta$) few hosts put out a significant fraction of the total incidence (\cref{fig:infectionsource:d}).
In other words, these hosts are responsible for the spread of the plasmid through the system, and all other hosts continuously get (re)infected from them (\cref{fig:infectionsource:b}).

Consequently, these species indeed act as \emph{sources} from which the plasmid can rapidly spread through the community when environmental conditions change, e.g. when genes on the plasmid instead confer fitness advantages (\cref{app:subsec:maintenanceimportance}).
Additionally, depending on the infection network, maintenance within these critical species is typically realized solely by continuous self-infection [as these rates are typically higher, see~\nameref{sec:methods} and recall, e.g., Refs.~\cite{tamminen2012largescale,hu2016bacterial,alderliesten2020effect}], or infection within a very small subcommunity of connected host species.
Note that such dynamics, more formally known as \emph{source-sink dynamics}, have been previously observed in model host-plasmid systems~\cite{hall2016source}, and thus our results suggest that these dynamics may additionally underlie plasmid maintenance in large communities as well.

To reiterate; in all cases above, purely from randomness may emerge a host --- or more specifically, a host-plasmid combination --- that meets just the right criteria that enables a plasmid to be maintained indefinitely, regardless of any other system properties.

\subsection{Effective network properties in Lotka-Volterra systems}%
Perhaps counterintuitively, the source-sink dynamics in the presence of super-hosts suggest that the network structure of the interaction and infection networks should, in fact, \emph{not} significantly alter plasmid maintenance.
The reason is that a single host is (or a select few hosts are) responsible for the plasmid's maintenance, and whether these species maintain the plasmid or not depends solely on the species themselves, and very little on (structured) interactions with other species.
Of course, plasmids are only able to infect those who are in the same connected component (recall~\cref{fig:infectionsource:b}), and the plasmid can be maintained only if the ecological dynamics do not reduce the size of these components quickly and significantly (\cref{fig:networkeffects:a}).
If they do, functional extinctions may ``break apart'' these networks, which subsequently may reduce maintenance probabilities.

\begin{figure}[t!]
  \centering
  \begin{tikzpicture}
    \node[anchor=west] (networks) {%
      \includegraphics[width=.775\columnwidth]{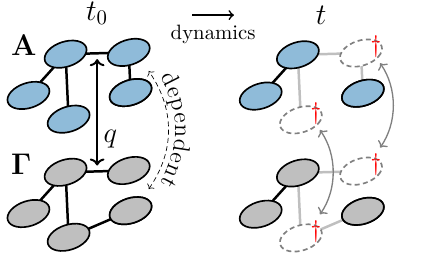}%
    };
    \node[anchor=north,xshift=-2em] (effdegree) at (networks.south) {%
      \includegraphics[width=.875\columnwidth]{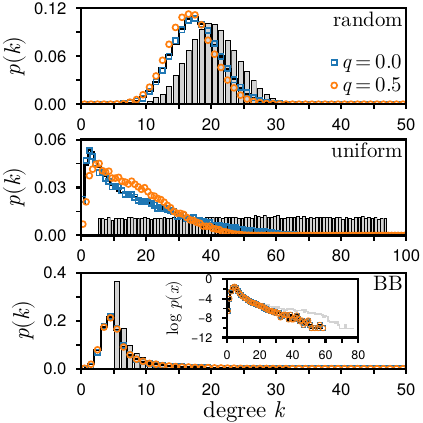}%
    };
    \node[anchor=center,inner sep=0pt] (b) at (effdegree.north west) {\textbf{b.}};
    \node[anchor=center,inner sep=0pt] (a) at (b |- networks.north west) {\textbf{a.}};
  \end{tikzpicture}
  \caption{%
    \textbf{Effective networks emerge from dynamics.}
    \textbf{a.}~A schematic illustration of the edge overlap probability $q$ that defines the probability that an edge that exists in $\mathbf{A}$ also appears in $\boldsymbol{\Gamma}$ at time $t_0$.
    In addition, it illustrates the effects of ecological dynamics such that at a time $t \gg t_0$ functional extinctions have drastically changed network properties.
    \textbf{b.}~Effective network properties for three types of networks: random (Erd\H{o}s-R\'enyi), networks with a uniform degree distribution, and a Bianconi-Barab\'asi (BB) fitness network with a uniform distribution for the fitness (see~\nameref{sec:methods} for more details).
    Grey bars depict the degree distribution of the interaction matrix $\mathbf{A}$ at the start of the simulation (i.e., at time $t_0$).
    The black line is the degree distribution of the interaction network after the system has reached a stable state and extinctions have occurred.
    Square and round markers depict the degree distribution for the infection network $\boldsymbol{\Gamma}$ for two overlap propabilities $q$.
    Note that densities are such that the networks still contain a large connected component.
    The lower inset indicates the same data but in a logarithmic scale, to illustrate the (near) power law behavior of the degree distribution (grey line).
  }
  \label{fig:networkeffects}
  \sublabel{fig:networkeffects:a}
  \sublabel{fig:networkeffects:b}
\end{figure}

To investigate whether extinction-driven changes to the effective interaction and infection networks changes plasmid maintenance probabilities, we considered a final adaptation that changes how the infection network $\boldsymbol{\Gamma}$ is sampled.
Briefly, once the interaction network $\mathbf{A}$ is sampled, we consider the infection network with the same edge density (number of edges), but with probability $q$ a random edge in $\mathbf{A}$ is maintained in $\boldsymbol{\Gamma}$ (\cref{fig:networkeffects:a}), and with probability $1-q$ its source and destination are chosen at random (while preserving the degree, \nameref{sec:methods}).
Such a model allows us to interpolate between the structurally independent case ($q=0$) and the structurally equal case ($q=1$).

Our results suggest that the distinct infection networks that we considered rarely break apart due to functional extinctions.
The main reason is that, while a perhaps an ecologically relevant portion of hosts functionally goes extinct ($\approx 10\%$ on average, \nameref{sec:methods}), the networks are initially too dense such that the largest connected component is unaffected by node removal, regardless of the value of $q$ (\cref{fig:networkeffects:b}).
We believe it is important to note that functional extinction \emph{does} alter the network properties [such as the degree distribution $p(k)$] significantly --- something which has lately been discussed in studies of the Lotka-Volterra model~\cite{aguirre-lopez2024heterogeneous,park2024incorporatinga,poley2024interaction}.
Yet, it is the size of the largest connected component that barely changes, and it is this feature of the network alone that underlies plasmid maintenance as the super-host that maintains the plasmid on its own will surely be infected.
While we could, of course, alter network properties or system parameters so as to increase the number of extinctions, the very fact that such restrictions are necessary to eradicate plasmids from these large systems strongly support their observed ubiquity in natural communities.
It furthermore suggests that extinction-driven rewiring of plasmid infection networks is unlikely to affect plasmid maintenance significantly.

\section*{Discussion}
\hypertarget{sec:discussion}{}  
In this work, we introduced a generic model that combines both ecological and epidemiological dynamics.
Its generic form allowed us to investigate how plasmid maintenance is affected by statistical patterns instead of host- or plasmid-specific parameters.
In doing so, we have shown that the ability of a community to maintain a focal plasmid depends on the distributions of species abundances and infection rates.
More specifically, our results demonstrate that sufficient heterogeneity in either of the distributions can ensure plasmids to be maintained indefinitely.
Our results furthermore suggest that plasmid maintenance --- and likely maintenance of similar mobile genetic elements as well --- becomes virtually inevitable when systems exhibit sufficient heterogeneity, even in the absence of positive selection.

The complexity of microbiomes is daunting, and predictive models of microbiome activity often depend on many difficult-to-obtain parameters.
This frustrates efforts to develop a general understanding of key processes in microbial communities.
The work presented here posits an alternative; easier-to-obtain distributions can critically inform our understanding of plasmid dynamics.
For example, in recent years there has been a rising interest in using plasmids as vehicles for disseminating functional traits into environmental microbiomes~\cite{garbisu2017plasmidmediated,carroll2018plasmid,french2020horizontal,marquiegui-alvaro2025genetic}.
These efforts are motivated by observations that plasmid-encoded enzymes can help degrade various pollutants, including herbicides, hydrocarbons, and plastic substrates~\cite{bhatt2021plasmidmediated,marquiegui-alvaro2025genetic}.
The advantage of plasmid-mediated degradation is that they can transfer traits into locally adapted microbiomes [a process also called \emph{bioaugmentation}~\cite{omokhagboradams2020bioremediation}], which avoids having to introduce novel species for which establishment is difficult.
To this end, assessing the potential of a community to maintain a new plasmid could enhance and inform such approaches.
Such assessments are additionally relevant to investigate or manage potential risks posed by undesirable traits --- most notably antibiotic resistance.
For example, profiling patient microbiomes for plasmid susceptibility may help to inform possible treatments~\cite{leon-sampedro2021pervasive}.
Similar assessments in an agricultural context could assist with preventing the establishment of environmental reservoirs of antimicrobial resistance~\cite{macedo2022horizontal}.

Although we can predict plasmid maintenance in model ecosystems, our approach is of course not without its limitations.
Perhaps the most notable limitation is the poor scaling with the number of plasmids~\cite{wang2020persistence}.
Natural systems contain a vast amount of different plasmids and the combinatorial problem of predicting which plasmids will be maintained quickly becomes intractable (\cref{sec:app:nmges}).
To accommodate this, novel, plasmid-centric models, are much needed~\cite{wang2020persistence,wang2021predicting,zhu2024horizontal}.
Additionally, whereas we have implicitly assumed that all members of the community can host a focal plasmid, it is likely that natural communities will contain members that cannot harbor it at all.
As such, future models should properly take plasmid-host compatibility into account~\cite{novick1987plasmid,dionisio2019interactions,zhu2024horizontal} (and see Ref.~\cite{wang2025interplay} for a recent attempt by the authors).
However, do note that host-plasmid compatibility, which defines the subcommunity wherein the plasmid can spread, could in principle be defined in advance from data on known plasmid host ranges, or could be determined experimentally for a given community using approaches such as fluorescence-activated cell sorting~\cite{klumper2015broad} or epicPCR~\cite{roman2021epicpcr} --- after which the analyses provided here may be applied.

Another limitation is that in order to predict plasmid maintenance, our model requires information about the distribution of conjugation rates, which we here assumed to be a log-normal distribution.
This assumption is consistent with reported measurements of conjugation rates between plasmid-susceptible hosts~\cite{quon2025quantifying}.
Still, estimating the distribution of infection ratios for real-world communities of interest is likely to be the greatest challenge for implementing our approach, but high(er) throughput methods for assessing conjugation rates in development will likely make such efforts more attainable~\cite{alalam2020highthroughput,arbe-carton2023developmenta}.

Next, we have used deterministic models instead of stochastic ones, and have let the randomness manifest itself in the parameters of the model and not in the dynamics themselves.
Dynamical stochasticity is, however, known to drastically alter pathogen spread in models of epidemics~\cite{lloyd-smith2005superspreading,penn2023intrinsic}, and can introduce ecological extinctions that are absent in deterministic models~\cite{dobrinevski2012extinction}.
While we acknowledge that a stochastic model of our generic framework is a valid path forward, we hypothesize that this would lead to patterns similar to the ones described here.
More specifically, as our results indicate that randomness dramatically impacts plasmid maintenance, even in a deterministic setting, we expect that the addition of stochasticity is likely to exacerbate these effects~\cite{paulsson2003plasmids}.

We have not considered other mechanisms that `solve' the plasmid paradox to explain maintenance without selection.
For example, it is known that plasmid costs can be ameliorated by host-plasmid co-evolution~\cite{dahlberg2003amelioration,wein2019emergence}, such that the host can take advantage of the genetic material without paying the normally associated fitness cost~\cite{harrison2012plasmidmediated,millan2014positive}.
Plasmids may also explicitly change species interactions~\cite{bergstrom2000natural,newbury2022fitness}, thus directly modulating the dynamics themselves.
As these dynamics underlie their maintenance, it is unclear whether such mechanisms are more important than conjugation-related mechanisms in diverse microbiomes.

Camacho-Mateu \emph{et al.}~\cite{camacho-mateu2024nonequilibrium} rightfully point scientists towards the growing tension between the ever-increasing amount of empirical data and the generic patterns that analytical and numerical investigations put forward.
While the amount of data is staggering~\cite{nayfach2021genomic}, its use for fitting generic models has generally not increased with the same rate --- something that becomes more apparent when the models themselves grow in complexity to account for additional mechanisms.
In fact, this seemingly parallels the increased awareness that abundance distributions are perhaps not a `\emph{catch-all}', and that measurements of microbial systems need to be more detailed in order to uncover the mechanisms that underlie their dynamics.
This could lead to an escalation in both model and data complexity, which we believe is unlikely to lead to an increased understanding of the important mechanisms at play.
These problems are similar to the main problem of \emph{testability} in microbiology, as most models for microbial dynamics are in practice very hard to test, or data is too sparse to appropriately fit a flexible model~\cite{servan2021tractable}.
To this end, novel tractable and testable models of microbial ecosystems are much warranted.

Despite its limitations, the work presented here emphasizes the seemingly ever increasing utility of metagenomic sampling of microbiomes.
The ongoing development of tools, such as the analyses presented here, continue to enable novel ways of exploiting the vast amount of available data.
In our case, our results demonstrated that knowing species abundance and plasmid conjugation rate distributions is sufficient to predict when plasmid maintenance is guaranteed.
In addition, we uncovered a simple and intuitive mechanism by which plasmids can be maintained in microbial communities: randomness.
That is, sufficiently random and heterogeneous systems are inherently more likely to maintain plasmids.
This urges us to rethink whether plasmids would ever truly go extinct in natural microbial communities, or whether their persistence is guaranteed if these communities are sufficiently diverse.

\matmethods{
  \label[Materials and methods]{sec:methods}%

  \subsection*{A generic model of population dynamics with mobile genetic elements}%
  We study a generic ecological-epidemiological model with $S$ distinct host species and $N$ mobile genetic elements (MGEs).
  Within the context of the model, each host can either host any combination of MGEs, or it can be free of any element.
  A subpopulation corresponds to a specific host $i$ hosting a particular combination of elements, labelled with $u$.
  The total number of \emph{subpopulations} equals $S_{\textrm{tot}} = S \cdot 2^N$.
  Inspired by similar models~[see, e.g., Refs.~\cite{holt1985infectious,vandendriessche2004disease,mccormack2006disease,hall2016source,venturino2016ecoepidemiology,lopatkin2017persistence,coyte2022horizontal}], we introduce a generic model here that considers the abundances of each subpopulation $x_i^u$ to be regulated by two mechanisms (see also~\cref{fig:scheme,fig:elementaryreactions} and \cref{app:sec:model});
  \begin{equation}
    \label{eq:methods:glvp}
    \odv{x_i^u}{t} =
    f_i^u(\boldsymbol{x}; \mathbf{A}, \boldsymbol{z}) + \gamma_i^u(\boldsymbol{x}; \boldsymbol{\Gamma},\boldsymbol{z})
  \end{equation}
  where \emph{ecological dynamics} (e.g., competition) are captured by $f_i$, and \emph{epidemiological dynamics} (e.g., infection) by $\gamma_i$.
  In addition, we assume that these functions depend on interaction and infection \emph{networks}, $\mathbf{A}$ and $\boldsymbol{\Gamma}$ respectively (see~\cref{fig:scheme}).
  We further consider both mechanisms to be affected by plasmid \emph{traits}, denoted with $\boldsymbol{z}$.
  For example, a common trait is that plasmids typically reduce the growth rates of their hosts by some amount~\cite{bergstrom2000natural,sorensen2005studying,lili2007persistence,summers2009biology}.

  Note that~\cref{eq:methods:glvp} is at its core a compartmental model wherein the total number of individuals of a specific host $i$ is \emph{not} fixed.
  As such, it readily captures well-known models with a proper choice of the ecological functions and transmission and background processes.

  \subsection*{Generalized Lotka-Volterra systems with interaction disorder}%
  While~\cref{eq:methods:glvp} is generally complicated to study, its ecological part obtained by disregarding infection-processes (i.e., $\gamma_i=0$) has recently received a lot of attention within the field of theoretical ecology (see Ref.~\cite{altieri2025unveiling} for a recent overview).
  Here, we assume the ecological dynamics to be of the Lotka-Volterra type with (constant) dilution rate $d$, which read
  \begin{equation}
    \label{eq:methods:glv}
    \odv{x_i}{t} \equiv f_i(\boldsymbol{x}; \mathbf{A}) = \frac{r_i x_i}{K_i}
    \bigg( K_i + \sum_{j} A_{ij} x_j \bigg) - d x_i,
  \end{equation}
  where we have now omitted the infection dynamics and have hence dropped the superscript for the pathogenic state $u$.
  Most notably, these models have been thoroughly investigated in the so-called \emph{disordered} limit~\cite{bunin2017ecological,barbier2021fingerprints,hu2022emergent,galla2024generatingfunctional,mallmin2024chaotic}.
  In this limit, one considers the $S(S-1)$ interaction coefficients that define the interaction matrix $\mathbf{A}$ to be drawn from a distribution, which is typically a Gaussian with some mean $\mu_A$ and variance $\sigma_A^2$.
  Interaction sparsity is determined by the \emph{connectance}~$c_A$, which defines the probability with which two species interact.
  Under these conditions, elements of the interaction matrix $\mathbf{A}$ are zero with probability $1-c_A$, and the non-zero elements are defined as
  \begin{equation}
    \label{eq:A}
    A_{ij} = \frac{\mu_A}{S} + \frac{\sigma_A}{\sqrt{S}} b_{ij}
  \end{equation}
  with $b_{ij} \sim \mathcal{N}(0,1)$ a standard normal random variable.
  Diagonal terms, often called \emph{self-interactions}, are taken equal to unity, $A_{ii}=1$.
  The reason for the (inverse) scaling with $S$ (or $\sqrt{S}$) is twofold.
  One is to ensure that the thermodynamics limit $S\rightarrow \infty$ remains meaningful when $S \rightarrow \infty$, and the other is to ensure stability of numerical integration schemes (see below).
  One can further allow for correlations between elements $A_{ij}$ and $A_{ji}$ to reflect patterns of interest, such as letting $\rho_A = \textrm{corr}(A_{ij} A_{ji})$.
  We assume $\rho_A = 0$ unless mentioned otherwise, but non-zero values are discussed in~\cref{subsec:app:correlations}.
  While there is currently much debate on the amount of mutualism vs. competition in microbial systems [see, e.g., Refs.~\cite{coyte2015ecology,kehe2021positive,palmer2022bacterial}], we assume that, on average, species compete ($\mu_A > 0$ and $\sigma_A$ such that mutualistic interactions are rare).

  \subsection*{Species abundance distributions}%
  Following some relatively mild constraints ($d = 0$, and see~\cref{app:sec:sads}), the assumption that systems are disordered (i.e., random interactions), allows us to write down a closed form solution of the species abundance distribution (SAD) of the steady state, which is one of the components that are necessary to predict element maintenance from a statistical perspective.
  When elements of the interaction matrix follow~\cref{eq:A}, the species abundance distribution of the steady state $\boldsymbol{x}^\star$ is a rectified Gaussian, which reads~\cite{bunin2017ecological,galla2024generatingfunctional}
  \begin{equation}
    \label{eq:rectgaussian}
    p(x^\star) = (1-\phi) \delta(x^\star) + p^+(x^\star) \Theta(x^\star)
  \end{equation}
  where $\phi$ is the fraction of species that survive the dynamics of~\cref{eq:glvp} (see the numerical implementation details below), $p^+$ the abundance distribution of the surviving species, $\Theta(x)$ the Heaviside step function which is $1$ when $x>0$ and $0$ otherwise, and where we have expressed the statistical equivalence of all species by letting $x^\star \sim x_i^\star$.
  In particular, $p^+$ is a Gaussian distribution with its moments depending solely on the properties of the interaction matrix.
  While empirical abundance distributions are most often not Gaussian (see main text), as they display heavy-tails~\cite{shoemaker2017macroecological,eguiluz2019scaling}, the expression for $p(x^\star)$, serves as a useful proxy when reasoning about possible factors that underly element maintenance and allows for comparison with numerical experiments.

  \subsection*{The next-generation matrix formalism}%
  To compute the basic reproduction number --- which effectively determines whether an MGE will be maintained --- we rely on established methods from epidemiology, most notably the next-generation matrix (\cref{app:sec:ngm}).
  Here, we briefly review the next-generation matrix formalism, which has been developed to investigate whether a ``disease'' (e.g., a pathogen, or an MGE such as a plasmid) will become endemic (i.e., it will remain indefinitely), and under what conditions it will be eradicated from the population~\cite{diekmann1990definition,vandendriessche2002reproduction,diekmann2009construction}.

  One writes their system of interest [\cref{eq:methods:glvp}, now with $\gamma_i \neq 0$] in such a way that one has a ``disease-free'' subsystem, here denoted with $\boldsymbol{x}$, and a system wherein the dynamics of the disease $\gamma_i$ is specified, i.e., the infected subsystem, denoted with $\boldsymbol{y}$.
  Then, the infected subsystem is written as
  \begin{equation}
    \odv{\boldsymbol{y}}{t} = \boldsymbol{g} - \boldsymbol{h}
  \end{equation}
  where $\boldsymbol{g}$ captures all the disease-related dynamics (such as compartment changes through infection, etc.), and $\boldsymbol{h}$ all other compartment changes (such as growth or death processes).
  We then write the Jacobian matrices $\mathbf{G}$ and $\mathbf{H}$ that are defined at the \emph{disease-free equilibrium} (DFE) for which $\boldsymbol{y}=0$.
  That is, for a generic pathogenic profile $u$ (see~\cref{eq:methods:glvp} and~\cref{app:sec:model}), we can write
  \begin{equation}
    \odv{y_i^u}{t} = g_i^u - h_i^u  
  \end{equation}
  which subsequently defines the Jacobians as
  \begin{equation}
    G_{ij}^{uv} = \left( \pdv{g_i^u}{y_j^v} \right)_{\boldsymbol{y}=0} \qquad \textrm{and} \qquad
    H_{ij}^{uv} = \left( \pdv{h_i^u}{y_j^v} \right)_{\boldsymbol{y}=0}
  \end{equation}
  While generally one can regard these Jacobians as tensors, a simple relabeling scheme is enough to flatten these tensors to matrices, which, with some abuse of notation, read
  \begin{equation}
    \label{eq:GH}
    G_{ij} = \left( \pdv{g_i}{y_j} \right)_{\boldsymbol{y}=0} \qquad \textrm{and} \qquad
    H_{ij} = \left( \pdv{h_i}{y_j} \right)_{\boldsymbol{y}=0}
  \end{equation}
  Here we have simply relabeled some combination of species $i$ with pathogenic profile $u$ to be indicated with subpopulation $i$, by recognizing that the label is only useful for some ordering of the subpopulations, from which the dynamics is independent.
  Then, the next-generation matrix $\mathbf{K}$ is defined as
  \begin{equation}
    \mathbf{K} = \mathbf{G} \mathbf{H}^{-1}  
  \end{equation}
  In compartment models, such as ours, this matrix defines the expected number of new infections produced by individuals in other compartments.
  Subsequently, the \emph{basic reproduction number} $R_0$, which is the spectral radius of $\mathbf{K}$, represents the number of secondary infections as a result of a single infected individual.
  In our case, $\mathbf{K}$ is a non-negative matrix, and one readily finds $R_0 = \max_i \textrm{Re} \lambda_i$, with $\textrm{Re}\lambda_i$ the real part of eigenvalue $\lambda_i$.
  Briefly, from the definitions, it follows that when $R_0>1$, the disease becomes endemic, while for $R_0 < 1$ the disease goes extinct.

  \paragraph{Plasmid conjugation}%
  To show that abundance distributions and knowledge about the infectious traits (e.g., conjugation rates) of a focal plasmid are sufficient for predicting its maintenance, we considered a relatively simple non-linear infection model akin to standard epidemiological ones.
  That is, we define
  \begin{equation}
    \label{eq:methods:gamma}
    \gamma_i(\boldsymbol{x}; \boldsymbol{\Gamma}, \boldsymbol{z}) =
    \sum_{j,v,w} \Gamma_{ij}^{uvw} x_i^v x_j^w + \sum_v \Omega_i^{uv} x_i^v,
  \end{equation}
  where $\Gamma_{ij}^{uvw}$ and $\Omega_i^{uv}$ are entries of the conjugation, or infection, tensor and plasmid segregation, or recovery tensor, respectively (see~\cref{fig:elementaryreactions} and~\cref{app:sec:ngm}).
  Note that the same relabeling scheme as in~\cref{eq:GH} can be applied, which is especially useful when a single focal plasmid is of interest.
  In that case, we denote with $x_i$ and $y_i$ the subpopulations that are free of the plasmid or hosting the plasmid, respectively.
  When the number of plasmids is more than one, elements of the infection tensor $\boldsymbol{\Gamma}$ capture transmision of an element \emph{within} profile $v$ to $w$ that (potentially) leads to a new element profile $u$ (see~\cref{fig:elementaryreactions}).
  While here we limit ourselves to the study of a single focal plasmid, the framework presented here is generic and any number of plasmids (or different MGEs) can, in principle, be chosen.
  However, we would like to mention that a combinatorial explosion in the number of subpopulations and parameters makes this generic model not suitable for studies with more than a handful of MGEs and other models, such as those in~Refs.~\cite{wang2020persistence,wang2021predicting,zhu2024horizontal} should be considered instead.

  \subsection*{Details on numerical integration of disordered Lotka-Volterra systems}%
  All numerical details on abundance distribution, maintenance probabilities, and population dynamics, have been obtained by numerically integrating~\cref{eq:glvp}, in conjuction with Lotka-Volterra dynamics of~\cref{eq:methods:glv} and infection dynamics as in~\cref{eq:methods:gamma}.
  We used \texttt{Julia} and have exploited fast, in-place solvers of (systems of) ordinary differential equations using and~\texttt{DifferentialEquations.jl}~\cite{rackauckas2017differentialequationsjl}.
  Unless mentioned otherwise, solutions have been obtained using Runge-Kutta pairs of order 5(4) with adaptive time stepping, as described in~\cite{tsitouras2011runge} with automated stiffness detection that switches to an order 2(3) Rosenbrock-W method (in \texttt{DifferentialEquations.jl}, this method is encoded under the name \texttt{AutoTsit5(Rosenbrock23())}).
  Note that while different solvers may give (slightly) different results depending on their (default) error tolerances, these should not alter the results presented here.

  We additionaly handle extinctions explicitly.
  As we cannot numerically integrate over infinite time windows, in the strict sense species will never go truly extinct (i.e., $x_i > 0$ for all $t < \infty$).
  To this end, we set abundances $x_i = 0$ when they fall below a chosen threshold $\vartheta$.
  We have found no differences between doing this during the integration or at the end.
  For numerical stability we chose the former: species abundances are set to exactly $0$ when $x_i < \vartheta = 10^{-9}$ at any time during the integration.

  To measure plasmid maintenance from these simulations, we first let the ecological dynamics without a plasmid [i.e., $y_i=0$ and $\gamma_i = 0$, see~\cref{eq:glvp}] converge until a time $t = 10^6$, after which we introduce a small amount of plasmids into a single host.
  More specifically, we let $y_i > 0$ (but $y_i \ll x_i$), and again numerically integrate the system but now with $\gamma_i$ as in~\cref{eq:methods:gamma}.
  We again let this system converge to a steady state by integrating again until $t = 10^6$.
  The abundances $\boldsymbol{x} = (x_1, \ldots, x_S, y_1, \ldots, y_S)$ are then collected for further analysis.
  For example, plasmid maintenance probabilities can be obtained simply by counting whether there exists a species for which $y_i > \vartheta$ for all realizations of the dynamics.
  Unless mentioned otherwise, all results are computed over 256 independent realizations of the dynamics.

  \subsection*{Data analysis of relative abundance distributions}%
  Data on relative abundances distributions have been obtained from Ref.~\cite{li2017gut}.
  An already parsed dataset has been made available by Grilli~\cite{grilli2020macroecological} (see the accompanied \href{https://github.com/jacopogrilli/lawsdiv}{repository}).
  The dataset contains relative abundances for each of the samples of a distinct experiment (a measurement).
  These abundances are used to compute the next-generation matrix by using~\cref{eq:Kij}.
  As our model is agnostic to the specific species, for each of the samples we obtain an array of relative abundances and we compute the next-generation matrix $\mathbf{K}_l$ for each of the samples.
  Then, for each sample we obtain a basic reproduction number $R_0^{(l)}$ and by aggregating all samples we obtain a distribution $p(R_0)$ which we use to estimate plasmid maintenance probabilities $P[R_0 > 1]$ (\cref{fig:R0gut}).

  \subsection*{Data analysis for distributions of infection rates}%
  To motivate our choice for the distribution over infection rates, we have used data on MGE conjugation rates in both clinical and environmental settings from Ref.~\cite{quon2025quantifying}.
  These data contain mean rates of horizontal gene transfer for a diverse set of MGEs and host species.
  By aggregating all species and environments and computing a histogram over conjugation rates, we find that the log-normal distribution is an excellent fit (\cref{app:sec:distributions}).
  Whereas a full statistical examination of conjugation rates in microbiomes is out of the scope of this work, it acts as empirical evidence that supports our assumption of log-normally distributed infection rates.
  Note that it is additionally known that plasmids conjugate more rapidly depending on their phylogenetic distances~\cite{tamminen2012largescale,hu2016bacterial}.
  To reflect this, when sampling conjugation rates, we instead let the rate of self-infection $\Gamma_{ii}$ to be 10 times the sampled value that would otherwise be considered~\cite{alderliesten2020effect}.

  \subsection*{Heterogeneous plasmid traits}%
  Whereas the above description has affirmed the log-normal distribution for infection ratios $\beta_{ij}$, we further assume the plasmid fitness costs $\alpha_i$ to be distributed according to a log-normal distribution.
  The variance, mediated by the parameter $\sigma_\alpha$, is chosen depending on the desired mean fitness cost $\bar{\alpha}$.
  This is done as to avoid $\alpha_i > 1$, as we are interested here in mechanisms underlying plasmid maintenance in the \emph{absence} of positive selection.
  More specifically, for a given $\bar{\alpha} \leq 1$ we define $\sigma_\alpha$ such that $\zeta = 0.997$ (99.7\%, i.e.~$\pm$3 standard deviations) of the samples from $p(\alpha_i)$ are within $\sigma_\alpha$ from $\bar{\alpha}$.
  To do so, we compute the corresponding $z$-value for the given confidence interval as $z = \sqrt{2} \cdot \textrm{erf}^{-1}(2\zeta - 1)$, and use it to compute the parameters of the log-normal distribution
  \begin{equation}
    \sigma_\alpha = \tfrac{1}{2} \left[ z - \sqrt{z^2 + 2\log(\bar{\alpha})} \right],
    \quad
    \mu_\alpha = -\frac{\sigma_\alpha^2}{2} + \log(\bar{\alpha})
  \end{equation}
  Then, $\alpha_i \sim \textrm{LogNormal}(\mu_\alpha, \sigma_\alpha)$, which has the desired mean $\bar{\alpha}$.
  We are aware that this procedure results in relatively small variances on $\alpha_i$, but relatively recent investigations have indicated that most fitness effects are slight (i.e., $\alpha_i \approx 0.9$) and for most host species they are negative $\alpha_i < 1$~\cite{alonso-delvalle2021variability}.
  Unless mentioned otherwise, we thus use $\bar{\alpha} = 0.9$ and essentially all costs are thus distributed between $0.8$ and $1.0$.
  Note that when plasmid costs are too high (low $\alpha_i$), numerical simulations reveal that plasmids get eradicated from the system, whereas positive selections ($\alpha_i > 1$) lead to trivial maintenance as the infected subpopulation simply grows faster than the uninfected one.
  We have found little to no effect of changes in the actual distribution of fitness effects on plasmid maintenance, other than the ones mentioned just now.

  \subsection*{Details on network ensembles}%
  Here we will briefly explain how to generate the networks that we have used in the manuscript.
  First, most of our results have been obtained by assuming networks to be random --- that is, Erd\H{o}s-R\`enyi networks.
  These networks are defined by the number of nodes $S$ and a probability $c$ of connecting two random nodes $i$ and $j$, the \emph{connectance}.
  These have been studied extensively for decades.
  What is worth mentioning here is that there exists a critical connectance $c^\prime = \log (S) / S$ such that, for large $S$, the network is almost surely connected when $c > c^\prime$.
  As our choice of $S=100$ is small (within the context of networks), one must typically choose $c$ slightly larger to ensure connected networks.
  The connectance, in the context of our model, is further related to the average number of interactions a species has (or the number of species it can infect), and hence this simple model provides a meaningful starting point.
  However, the degree distribution is approximately a Poisson distribution about the mean $cS$, for which it is unsure whether this is realistic.

  Other works have indicated that most species interact with very few others, while few interact with many~\cite{sole2001complexity}.
  Interestingly, when one assumes a uniform degree distribution, $p(k) \sim \textrm{Unif}(k_{\min},k_{\max})$, and uses the Chung-Lu configuration model~\cite{chung2002average}, the effective degree distribution that emerges from the dynamics follows this exact pattern [see also~Ref.~\cite{poley2024interaction}].
  As the resulting distribution (that is; the \emph{effective} network ensemble) is difficult to sample from directly (as many sampled degree sequences are not graphical), we instead consider the effective network when applicable.
  In our simulations, unless mentioned otherwise, we select $k_{\min}=5$ and $k_{\max} = S-5$.
  Note that for (relatively) small $S$, there may be fewer nodes with high degree than expected.

  Finally, extending the networks above, we consider a family of networks where a few nodes (the hubs) have a number of edges that drastically exceeds the expected number of edges.
  We assume networks for which the degree distribution follows a power-law.
  These networks are called \emph{scale-free networks}~\cite{newman2018networks}.
  We focus here on a specific type of scale-free networks; Bianconi-Barabasi networks~\cite{bianconi2001competition,bianconi2001boseeinstein}.
  These networks are growing networks that combine preferential attachment with a fitness scheme.
  Briefly, one starts with a few nodes and one-by-one adds nodes.
  Each step, $m$ edges are added between the newly introduced node and established nodes with probability
  \begin{equation}
    \Pi_i = \frac{\eta_i k_i}{\sum_j \eta_j k_j},
  \end{equation}
  where the sum goes over the number of nodes already in the network.
  Note that we do not allow self- or double-edges.
  It turns out that networks grown with this scheme have power-law degree distribution, with the exponent depending on the fitnesses $\eta_i$.
  For $\eta_i = \textrm{const.}$ one recovers the well-known Barabasi-Albert model.
  We choose instead a uniform distribution $\eta_i \sim \textrm{Unif}(0,1)$, and the resulting degree distribution turns out to have exponent $\approx 2.25$ in the limit of $S \rightarrow \infty$.
  As in our case $S$ is finite, we observe exponential truncations towards the upper limit, as seen in~\cref{fig:networkeffects:b}.
  For our purposes, however, it is important to note that for $m > 2$ the resulting network is dense, and even extinction of the hubs does not seem to disconnect the network and the giant component persists.
  This density effect also applies to the random and configuration model networks described above.

  \subsection*{Sampling networks with overlap}%
  We have sampled interaction and infection networks $\mathbf{A}$ and $\boldsymbol{\Gamma}$ with an \emph{overlap} $q$.
  The overlap $q$ is defined as the probability that an edge in $\mathbf{A}$ is also present in $\boldsymbol{\Gamma}$.
  More formally, let the total number of edges in $\mathbf{A}$ be $E$, and we now sample $\boldsymbol{\Gamma}$ with the same number of edges.
  For each edge $e_i$ in $\mathbf{A}$, with probability $q$ it is included in $\boldsymbol{\Gamma}$ by storing it in some array $\boldsymbol{e} = (e_1, e_2, \ldots)$, but with probability $1-q$ it is stored in a distinct array $\boldsymbol{e}^\prime$.
  After having considered all edges in $\mathbf{A}$, those who were not selected and put directly in $\boldsymbol{\Gamma}$ are now shuffled randomly while preserving the degree distribution.
  To do so, we select two edges from $\boldsymbol{e}^\prime$ randomly and swap their destinations such that the degree is preserved.
  This is done many times to effectively randomize the infection network, apart from the $qE$ edges that are kept.
  When $q=0$, all edges are shuffled and the infection network's topology is independent of the topology of the interaction network (but note that the number of edges is the same).
  Instead, when $q=1$ both networks are structurally identical as they have identical edges.

  \subsection*{Code availability}%
  Code to generate results and figures will be made available after acceptance.

}%
\showmatmethods{} 



\bibliography{bibliography.bib}

\clearpage
\makeatletter\@input{appendix-arxiv.tex}\makeatother


\section*{Supplementary material}

**Supporting Information Text**

## 1. A generic model of systems with transmissible elements

We start by introducing the elementary reactions that underlie our general eco-epidemic model. We want to emphasize that similar models exist (see, e.g., [1–7], among others), however that the particular way in which we capture distinct processes allows for a more general treatment. The elementary reactions we consider generally describe compartmental changes and refer to population and transmission dynamics. We consider a system of $S$ interacting species and $N$ transmissible elements. Generally, individuals (and hence subpopulations) can host any attainable combination of elements, and thus the maximum total number of subpopulations is $S_{tot} = S \cdot 2^N$. Then, we consider $X_i^u$ *individuals* of the host species $i$ that have pathogenic profile $u$. In other words, for all the $N$ pathogens, one can construct a binary vector that indicates whether a subpopulation hosts (is infected by) a particular pathogen. Each of the unique $2^N$ vectors is then labeled with $u$. We then assume the following elementary reactions

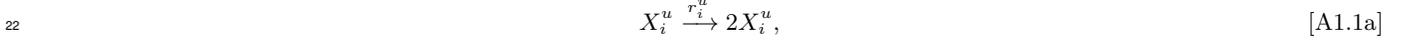
$$X_i^u \xrightarrow{r_i^u} 2X_i^u,$$ [A1.1a]

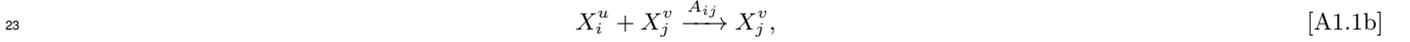
$$X_i^u + X_j^v \xrightarrow{A_{ij}} X_j^v,$$ [A1.1b]

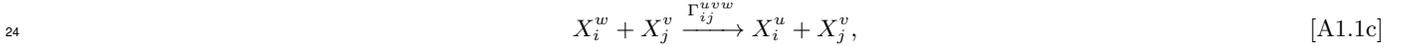
$$X_i^u + X_j^v \xrightarrow{\Gamma_{ij}^{uvw}} X_i^u + X_j^v,$$ [A1.1c]

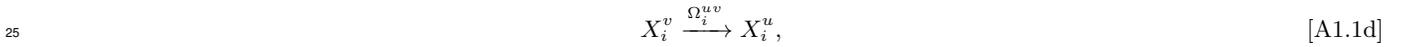
$$X_i^v \xrightarrow{\Omega_i^{uv}} X_i^u,$$ [A1.1d]

where from top to bottom we have growth with rate $r_i^u$, (competitive, see below) interaction with rate $A_{ij}$, infection with rate $\Gamma_{ij}^{uvw}$ and segregation (or recovery, or loss) with rate $\Omega_i^{uv}$. These reactions specify the form of our generic ecological-epidemiological model as a system of coupled ODEs of the abundance $x_i^u = X_i^u/C_i$, which we assume to be

$$\frac{dx_i^u}{dt} = x_i^u f_i(\boldsymbol{x}; \mathbf{A}, \boldsymbol{z}) + \gamma_i(\boldsymbol{x}; \boldsymbol{\Gamma}, \boldsymbol{\Omega}, \boldsymbol{z})$$ [A1.2]

where $f_i(\boldsymbol{x}; \mathbf{A}, \boldsymbol{z})$ a function that defines intra- and interspecific interactions captured by the interaction matrix $\mathbf{A}$, with $C_i$ the carrying capacity of host species $i$. Note that this function generally depends on the full state $\boldsymbol{x} = \{x_i^u \mid i = 1, \dots, S, u = 1, \dots, 2^N\}$ and some traits of the transmissible element $\boldsymbol{z}$ (e.g., infection rate). The second term pertains infection-related terms. As the elemental reactions dictate, we use a generic SIS-like model and define

$$\gamma_i(\boldsymbol{x}; \boldsymbol{\Gamma}, \boldsymbol{\Omega}, \boldsymbol{z}) = \sum_{j,v,w} \Gamma_{ij}^{uvw} x_i^u x_j^w + \sum_v \Omega_i^{uv} x_i^v,$$ [A1.3]

Elements of the *conjugation* (or infection) tensor $\boldsymbol{\Gamma}$ capture the transmission of an element within $v$ to $u$, that leads to the new element profile $w$ (again, see Fig. 2). Similarly, background rates (dilution, death, segregation, etc.) are captured by the *background tensor* $\boldsymbol{\Omega}$. These rates capture, for example, processes of a host losing one (or more) of its elements that leads to the compartmental change $v$ to $u$ (from which one may consider $\Omega_i^{uv}$ the recovery rate, as in more standard epidemiological models).

While seemingly complex, the above model simplifies significantly when one takes into account the fact that both the transmission and background tensor are typically sparse. The reason for that is that specific profiles changes simply cannot occur. For example, a species with a specific pathogenic profile can not simultaneously lose and gain a pathogen. Additionally, what one may consider is that there are symmetries in the tensor that can drastically simplify the above formulation. In particular, we can define the subpopulations that do not host any pathogen as $x_i^\emptyset$ and write the dynamics for both the element-free subpopulation, and the remaining subpopulations as

$$\frac{dx_i^\emptyset}{dt} = x_i^\emptyset f_i(\boldsymbol{x}) - x_i^\emptyset \sum_{j,v} \widetilde{\Gamma}_{ij}^v x_j^v + \sum_v \Omega_i^v x_i^v$$ [A1.4a]

$$\frac{dx_i^u}{dt} = x_i^u f_i(\boldsymbol{x}) + \sum_{j,v,w} \Gamma_{ij}^{uvw} x_i^u x_j^w - \sum_v \Omega_i^v x_i^v$$ [A1.4b]

where we have defined $\widetilde{\Gamma}_{ij}^v \equiv \Gamma_{ij}^{\emptyset\emptyset v}$, and where we have made explicit the symmetries in both the transmission and background tensors. While this formulation perhaps does not appear simpler, it highlights that the dynamics are effectively split into two distinct counterparts: one group of subpopulations that refers to the species that do not host any element $x_i^\emptyset$ (within the context of epidemiology, this is often referred to as the *disease-free population*), and one group of subpopulations that hosts any attainable combination of elements $x_i^u$, $u \neq \emptyset$. We will denote these two subpopulations with $x_i$ and $y_i^u$, to make the distinct difference between the two more apparent and clear. When one restricts the number of combinations, e.g., by restricting co-occurrence of specific elements within the same host, the effective number of species $S_{tot}$ can decrease drastically and the equations can become amendable to analytic or numerical treatments.

Within this context, of particular interest are the cases where there is only one transmissible element, and the case where no elements co-occur within the same host. Both cases are amendable to analytical derivations, yet the former moreso and hence we shall mostly focus on systems with a single transmissible element in this work. A more thorough analytical investigation of the general model is considered to be out of the scope of this work (but see Appendix 6).

 Johannes Nauta, Kaitlin A. Schaal, Ying-Jie Wang, James P.J. Hall, Shai Pilosof, Manlio De Domenico

## 2. The next-generation matrix

To predict whether a transmissible element will be maintained or not, we rely on computing basic reproduction numbers using the next-generation matrix $\mathbf{K}$ (4, 8, 9). In order to derive this matrix, one typically introduces the "infected" subsystem $\boldsymbol{y}$ (as above in Eq. (A1.4), and see Appendix 3 below) and the vectors $\boldsymbol{g}$ and $\boldsymbol{h}$, such that

$$\frac{\mathrm{d}\boldsymbol{y}}{\mathrm{d}t} = \boldsymbol{g} - \boldsymbol{h}, \qquad \text{or equivalently,} \qquad \frac{\mathrm{d}y_i^u}{\mathrm{d}t} = g_i^u - h_i^u \quad \text{for all } i, u \tag{A2.5}$$

Here, the infected subsystem is the system that contains all subpopulations that are hosting any number of transmissible elements — i.e., all subpopulations that are infected by some element(s), $\{y_i^u\}$. Then the Jacobian matrices $\mathbf{G}$ and $\mathbf{H}$ are defined as the partial derivatives at the element-$u$-free steady state $\boldsymbol{y}^u = 0$ ($y_i^u = 0$ for all $i$), with their elements defined as

$$G_{ij}^{uv} = \left(\frac{\partial g_j^u}{\partial y_j^v}\right)_{\boldsymbol{y}^u = 0} \quad \text{and} \quad H_{ij}^{uv} = \left(\frac{\partial h_j^u}{\partial y_j^v}\right)_{\boldsymbol{y}^u = 0} \tag{A2.6}$$

Subsequently, the next-generation matrix is defined as

$$\mathbf{K} = \mathbf{G}\mathbf{H}^{-1} \tag{A2.7}$$

This matrix defines the set of basic reproduction numbers $R_0^u$, which are given by the spectral radius of the next generation matrix constructed by considering each system free of elements $u$. These numbers are defined as the average number of infected individuals that are generated by introducing a small fraction of infected individuals in a population (4, 8–12). When any of the $R_0^u > 1$, an epidemic may occur wherein element $u$ — but not necessarily *only* element $u$ — are maintained indefinitely in the system. As such, when one can predict the values of the basic reproduction numbers, one can reason about the maintenance of transmissible elements. For a single focal element it *exactly* predicts whether it is maintained or not, and hence we shall focus on this case below.

## 3. Basic reproduction numbers of large, random systems

Now that we have introduced the basic reproduction number, and how it can be computed using the next-generation matrix, let us return to large systems. Again, we focus here on systems with a single focal element. The reason for this, as shall be highlighted below, is that this system, and the one where no elements be co-hosted, are amendable to analytical calculations, thus providing us with much information on the constraints under which an element will be maintained or not. Extensions to multiple competing plasmids are discussed in Appendix 6.

When there is only a single transmissible element ($N = 1$), the split into distinct categories mentioned above is most apparent. We write the hosts without and with the element as $x_i$ and $y_i$, and write the full dynamics as

$$\frac{\mathrm{d}x_i}{\mathrm{d}t} = x_i f_i(\boldsymbol{x}, \boldsymbol{y}) - x_i \sum_j \Gamma_{ij} y_j + \sum_j \Omega_{ij} y_j \tag{A3.8a}$$

$$\frac{\mathrm{d}y_i}{\mathrm{d}t} = y_i f_i(\boldsymbol{x}, \boldsymbol{y}) + x_i \sum_j \Gamma_{ij} y_j - \sum_j \Omega_{ij} y_j \tag{A3.8b}$$

Then, using the next-generation matrix formalism as described above, we construct vectors $\boldsymbol{g}$ and $\boldsymbol{h}$ for which $\boldsymbol{y} = \boldsymbol{g} - \boldsymbol{h}$, i.e.;

$$g_i = x_i \sum_j \Gamma_{ij} y_j, \qquad h_i = \sum_j \Omega_{ij} y_j - y_i f_i(\boldsymbol{x}, \boldsymbol{y}) \tag{A3.9}$$

Then, the Jacobian matrices read

$$G_{ij} = \frac{\partial g_i}{\partial y_j} = \Gamma_{ij} x_i^\star, \tag{A3.10a}$$

$$H_{ij} = \frac{\partial h_i}{\partial y_j} = \Omega_{ij} - \delta_{ij} \left[f_i(\boldsymbol{x}, \boldsymbol{y})\right]_{\boldsymbol{y}=0} = \Omega_{ij} - \delta_{ij} f_i(\boldsymbol{x}) \tag{A3.10b}$$

When $f_i(\boldsymbol{x})$ is an appropriate ecological model, the second term on the right-hand side of Eq. (A3.10b) vanishes at the element-free steady state $\boldsymbol{x}^\star$ by definition, as we find that $f_i(\boldsymbol{x}) = 0$ for all $i^\star$. When there is only one transmissible element, $\Omega_{ij} \equiv \Omega_{ij}\delta_{ij}$, hence the background tensor $\boldsymbol{\Omega}$ is a diagonal matrix, which is easily inverted to obtain

$$\mathbf{K} = \mathbf{G}\mathbf{H}^{-1} \equiv \mathbf{G}\boldsymbol{\Omega}^{-1} \tag{A3.11}$$

where $(\boldsymbol{\Omega}^{-1})_{ii} = 1/\Omega_{ii}$. We can analytically compute the eigenvalue spectrum, and as $\mathbf{K}$ is by construction a non-negative matrix this, by the Perron-Frobenius theorem, is equal to the largest real eigenvalue, which in turn defines the basic reproduction

---

*To see why $x_i^\star = 0$ does not give problems, it is enough to notice that for any reasonable ecological function one must also have $x_i^\star = 0$ to be a fixed point of the dynamics.

number $R_0 = \max_i \lambda_i(\mathbf{K})$. The eigenvalues depend on the species abundance distribution $x_i^\star \sim p(x)$, which depends on the ecological model $f_i(\boldsymbol{x})$.

Here we deem it important to notice that regardless of the structure of the transmission and background matrices, the eigenvalues of $\mathbf{K}$ depend strongly on the abundances of the element-free system $\boldsymbol{x}^\star$. This result is pivotal, as it indicates that one needs to study the abundance distribution of the system *without* any transmissible elements, and then insert the transmission and background processes of the elements *in isolation*. This information is enough to predict whether the element will be maintained or not.

**3A. Fixed infection and background rates.** Let us first assume fixed background and transmission rates, i.e., we let $\Gamma_{ij} = \Gamma$ and $\Omega_{ij} = \Omega$ for all $i, j$. In that case, the next-generation matrix of Eq. (A3.11) is a rank-one matrix, which has $S - 1$ eigenvalues which are zero, and a single non-zero eigenvalue equal to the sum of one of its columns. As the basic reproduction number is the right-most largest eigenvalue, we find that

$$R_0 = \beta \sum_i x_i^\star \tag{A3.12}$$

where we define $\beta = \Gamma/\Omega$ as the *infection ratio*. From this formulation, it immediately becomes even more obvious that element maintenance depends on the species abundance distribution. For example, for Lotka-Volterra systems (see Appendix 4A), these abundances are described by a rectified Gaussian, which consists of a $\delta$-peak at 0 that contains all species that went extinct, and a Gaussian distribution for those with positive abundances (13, 14). Within this context, because $x_i^\star$ are random variables who follow the rectified Gaussian distribution, $R_0$ becomes a random variate as well. For large $S$, the distribution of $R_0$ is a Gaussian distribution with a mean and variance equal to (15)

$$\mu_R = S\beta\mu_x, \qquad \sigma_R^2 = S\beta^2\sigma_x^2, \tag{A3.13}$$

which means that the epidemiological requirement for the element to become endemic — equivalent to $R_0 > 1$ in the deterministic case — is now given by the *probability* of $R_0 > 1$ under the assumption that $x_i^\star$ is a rectified Gaussian. As $R_0$ is Gaussian, this probability is easily written down and reads

$$P[R_0 > 1] = 1 - \Phi\left(\frac{1 - \mu_R}{\sigma_R}\right) \tag{A3.14}$$

with

$$\Phi(x) = \frac{1}{2}\left(1 + \operatorname{erf}\left(\frac{x}{\sqrt{2}}\right)\right) \tag{A3.15}$$

and $\operatorname{erf}(x)$ the error function. When computing $\mu_x = \langle x \rangle$ and $\sigma_x^2 = \langle x^2 \rangle - \langle x \rangle^2$ for the particular ecological dynamics of interest (e.g., Lotka-Volterra dynamics, see Appendix 4A), and inserting those in the distribution of Eq. (A3.14), we obtain the closed-form solution which is plotted in Fig. 3 in the main text.

**3B. Random infection and background rates.** If instead the background and transmission rates are random, the picture changes slightly. Motivated by recent observations [(16), Appendix 5], and because rates need to be positive by definition, we let both transmission and segregation rates follow a log-normal distribution (see Ref. (16), and Appendix 5). That is to say,

$$\log \Gamma_{ij} \sim \mathcal{N}(\mu_\Gamma, \sigma_\Gamma^2), \quad \text{and} \quad \log \Omega_{ii} \sim \mathcal{N}(\mu_\Omega, \sigma_\Omega^2), \tag{A3.16}$$

where $\mu_{\Gamma,\Omega}$ and $\sigma_{\Gamma,\Omega}^2$ the mean and the variance, respectively. In this case, we find that the infection ratio $\beta$ is also log-normally distributed, as

$$\beta_{ij} = \frac{\Gamma_{ij}}{\Omega_{ij}} \quad \Rightarrow \quad \log \beta_{ij} \sim \mathcal{N}(\mu_\Gamma - \mu_\Omega, \sigma_\Gamma^2 + \sigma_\Omega^2) \tag{A3.17}$$

Now recall the basic reproduction number is given by spectral radius of the next-generation matrix, which in this case is a random matrix as

$$\mathbf{K} = \mathbf{G}\mathbf{H}^{-1}, \tag{A3.18}$$

with $\mathbf{H}$ again a diagonal matrix, as above, and $\mathbf{G}$ now a matrix with elements

$$G_{ij} = \Gamma_{ij} x_i^\star \tag{A3.19}$$

Again, we fix $\Omega_{ij} = \Omega$, such that we effectively choose the distribution of $\beta_{ij}$ instead. Regardless of the distribution of the interaction ratios $p(\beta_{ij})$ — assuming that all moments exist and are finite — we can now write the next-generation matrix as

$$\mathbf{K} = \boldsymbol{\beta} \circ (\boldsymbol{u}\dot{\boldsymbol{x}}^T), \tag{A3.20}$$

where $\boldsymbol{u}$ the vector of all ones, and where we have used $\dot{\boldsymbol{x}}^T = (\boldsymbol{x}^\star)^T$ the transpose of $\boldsymbol{x}^\star$. Here, $\boldsymbol{\beta}$ is the matrix with elements $\beta_{ij}$. Hence, the task at hand is to compute, or approximate, the eigenvalue of $\mathbf{K}$ with the largest positive real part — or, more specifically, the largest positive real eigenvalue. Notice that Eq. (A3.20) can be written as

$$\mathbf{K} = \boldsymbol{\beta}\operatorname{Diag}(\boldsymbol{x}^\star), \tag{A3.21}$$

Johannes Nauta, Kaitlin A. Schaal, Ying-Jie Wang, James P.J. Hall, Shai Pilosof, Manlio De Domenico

which, for a given $\boldsymbol{x}^\star$ allows us to estimate the largest eigenvalue $R_0$. More specifically, for a *given* $\boldsymbol{x}^\star$, we employ a mean-field approximation and construct, with some slight abuse of notation, the 'average' matrix $\langle \mathbf{K} \rangle$ as

$$\langle \mathbf{K} \rangle \equiv \langle \boldsymbol{\beta} \rangle \, \text{Diag}(\boldsymbol{x}^\star) \tag{A3.22}$$

Now consider that *elements* of the 'average' matrix $\langle \boldsymbol{\beta} \rangle$ are simply $\bar{\beta}$, with $\bar{\beta}$ the mean of the distribution $p(\beta_{ij})$, and we readily obtain that

$$\langle K \rangle \approx \bar{\beta} \, \text{Diag}(\boldsymbol{x}^\star) \tag{A3.23}$$

This matrix is essentially the same as the same as the matrix above when we assumed homogeneous infection, but with the transformation $\beta \to \bar{\beta}$, from which the mean-field approximation become apparent. That is, the largest eigenvalue, and thus the basic reproduction number, is well-approximated by

$$R_0 \approx \bar{\beta} \sum_i x_i^\star \tag{A3.24}$$

Indeed, numerical verification shows that the basic reproduction number is indeed close to the above approximation (see Fig. 3 in the main text). While an approximation of the basic reproduction number by itself does not give us information about its distribution, it does highlight how this can be approximated, as from Eq. (A3.24) it follows that the distribution of $R_0$ likely follows from that of both $\beta_{ij}$ and $x_i^\star$.

To reiterate, the equation for the basic reproduction number $R_0$ above again highlights the affect of the abundance distribution on plasmid maintenance probabilities. Therefore, we can reason about plasmid maintenance by computing or observing abundance distributions, which we shall do in more detail below.

### 3C. On the importance of maintenance probabilities over maintenance abundances.

Before investigating the effects of the abundance distributions on plasmid maintenance, let us first consider whether computing the probability of maintenance is sufficient. We argue that in most cases, the actual abundance of infected hosts is not as important as knowing the conditions under which there is a non-zero fraction present in the system. To illustrate this, let us initially assume a very simple system with only a single host and a single pathogen. We let the element-free host population and its infected counterpart be $x$ and $y$, and consider the element to infect with rate $\gamma$ and be lost with rate $\omega$. In this case, assuming some non-zero dilution rate $d > 0$ and a fitness effect $\alpha$ (see Appendix 4A.2 below), the maintenance criterion on the infection rate can be derived relatively easily by computing the basic reproduction number as described above (Appendix 2). This condition on the (now a scalar) infection rate $\gamma$ reads

$$\gamma > \gamma_c, \qquad \text{where} \qquad \gamma_c = \frac{\omega r + (1 - \alpha) r d}{C(r - d)} \tag{A3.25}$$

with $\omega$ the rate at which the transmissible element is lost, as before.

To see why the maintenance probability often suffices, let us consider a simple experiment in the parameter region where $\gamma \gtrsim \gamma_c$. In this region, the virulent element with cost $\alpha$ persists indefinitely, yet its relative abundance is low — i.e., $y \ll x$. Now consider a perturbation occurring at some (large) time $t'$, and the effect of this perturbation reduces the growth rate of the host species. However, similar to empirical observations of plasmids (e.g., [17–20]), the transmissible element provides resistance against the perturbation, and hence hosts that carry the element still grow as if the perturbation was not present. In this case, we see that in a short time period after the perturbation is applied, the transmissible element quickly sweeps through the population, and nearly all hosts will be infected (Fig. S1a). These results extend to systems with more species (Fig. S1b). This simple example shows that maintenance probabilities are often sufficient when one anticipates potential perturbations to which the transmissible element provides resistance. Perhaps obviously, when carrying the element provides no benefit, its relative abundance will remain small.

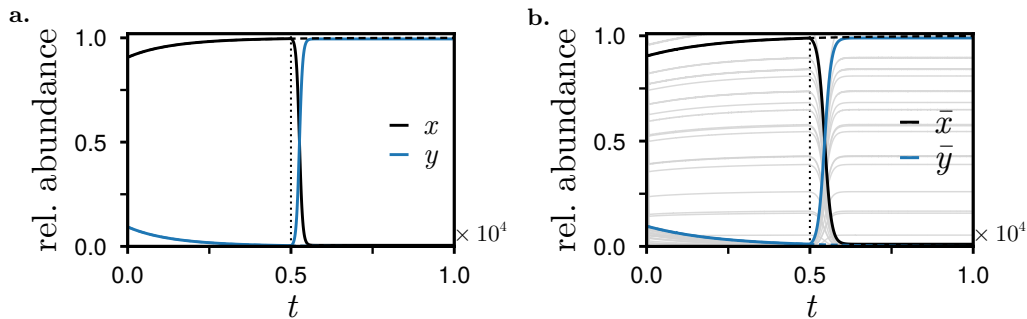

**Fig. S1.** Maintenance probabilities are sufficient as they predict potential sweeps under perturbations. **a.** At $t' = 5 \cdot 10^3$ (dotted vertical line) a perturbation is applied which reduces the growth rate of the single host species ($S = 1$), yet the plasmid ($N = 1$) provides resistance against this perturbation. After the perturbation, the infected subpopulation $y$ quickly takes over, as indicated by the relative abundance $x/(x + y)$ and $y/(x + y)$. These results have been obtained for $\gamma > \gamma_c$, yet with $x \gg y$ (see text). Dashed lines indicate population dynamics if carrying the plasmid did not provide any resistance. **b.** The same for a population of $S = 20$ species. Here, both the host abundances are shown (light gray) and the relative mean abundance $\bar{x} = \langle x \rangle / \sum_i (x_i + y_i)$ are shown. Grey lines indicate absolute abundances $x_i$ and $y_i$. Relevant parameters are $r = 1$, $C = 1$, $d = 0.05$, $\omega = 10^{-4}$, $\alpha = 0.8$, and $\gamma = 10^{-2}$ and $\gamma = 5 \cdot 10^{-4}$ for the $S = 1$ and $S = 20$ communities respectively.

Johannes Nauta, Kaitlin A. Schaal, Ying-Jie Wang, James P.J. Hall, Shai Pilosof, Manlio De Domenico

## 4. Effect of abundance distributions on plasmid maintenance

Let us now discuss in more detail the effects of the abundance distributions on plasmid maintenance. While one could measure these empirically and resort to a sampling scheme to investigate the effects of epidemiological parameters (as we have done, see main text and Appendix 5), for some systems, such as the generalized Lotka-Volterra system, the species abundance distribution can be derived exactly.

**4A. Generalized Lotka-Volterra systems.** When the population dynamics follows the generalized Lotka-Volterra model, expressions of the abundance distributions can be obtained. Using the cavity method, it has been shown that these distributions tend to a rectified Gaussian [13]. The same can be accomplished using generating functionals [14]. For completeness, we shall include here a brief overview of the distribution and how one may compute its moments, and shall omit a (lengthy) derivations, as those have been discussed elsewhere in detail [such as in Refs. [13, 14]]. The generalized Lotka-Volterra model reads

$$\frac{dx_i}{dt} = \frac{r_i x_i}{C_i} \left( C_i - x_i - \sum_{j \neq i} A_{ij} x_j \right) \qquad [A4.26]$$

To proceed, one assumes a fixed point ansatz — i.e., that the system reaches a steady state[†] — and, for $r_i = 1$ and $C_i = 1$, a rectified normal distribution can be derived that defines these fixed point abundances as

$$x^\star(z) = \max \left( 0, \frac{1 - \mu_A M_1 + z \sigma_A \sqrt{M_2}}{1 - \rho_A \chi \sigma_A^2} \right) \qquad [A4.27]$$

where $\chi$ can be considered as a susceptibility, and $M_{1,2}$ are the moments of the distribution $p(x_i^\star)$,

$$M_1 = \lim_{S \to \infty} \frac{1}{S} \sum_i \langle x_i^\star \rangle, \qquad M_2 = \lim_{S \to \infty} \frac{1}{S} \sum_i \langle (x_i^\star)^2 \rangle, \qquad [A4.28]$$

and $\mu_A$, $\sigma_A$, and $\rho_A$ the mean, standard deviation, and correlation of the interaction matrix **A**, respectively. For this, recall [Eq. [11]] that we assume $S$ large and the Lotka-Volterra model to be of the disordered type [13, 22–25], such that, for $i \neq j$, we have $A_{ij} = \mu_A/S + \sigma_A b_{ij}/\sqrt{S}$, with $b_{ij}$ correlated normally distributed variables with mean 0, variance 1 and correlation $\rho_A = \text{Corr}(A_{ij} A_{ji}) \equiv \text{Corr}(b_{ij} b_{ji})$. Note that the moments $M_{1,2}$ are moments of the unrectified distribution, which in reality includes species that have gone extinct.

In order to compute these moments, we require a set of self-consistent equations. It is helpful to introduce

$$\Delta = \frac{1 + \mu_A M_1}{\sigma_A \sqrt{M_2}} \qquad [A4.29]$$

and the integral

$$w_k(\Delta) = \frac{1}{\sqrt{2\pi}} \int_\infty^\Delta dq \, e^{\frac{1}{2} q^2} (\Delta - q)^k \qquad [A4.30]$$

Here, we follow [14], who finds

$$\sigma_A^2 = \frac{w_2}{(w_2 + \rho_A w_0)^2} \qquad [A4.31a]$$

$$\frac{1}{M_1} = \frac{\Delta}{w_1} \frac{w_2}{w_2 + \rho_A w_0} - \mu_A \qquad [A4.31b]$$

$$M_2 = \left( \frac{w_2}{w_2 + \rho_A w_0} \frac{M_1}{\sigma_A w_1} \right) \qquad [A4.31c]$$

$$\chi = w_0 + \rho_A \frac{w_0^2}{w_2} \qquad [A4.31d]$$

where, for brevity, we omitted the dependence on $\Delta$ for the $w_k$ functions. While this system can be solved using non-linear solvers, it turns out that these solutions depend on the initial guess, and also appear to be relatively unstable. To this end, a parametric solution can be obtained instead. This is done by fixing $\mu_A$ and $\rho_A$, but not $\sigma_A$. Then, using $\Delta$ as a free parameter — i.e., in the range of $-10$ to $10$, we can derive values of the other parameters as a function of $\Delta$ and $\sigma_A$. Then, one chooses the value of $\Delta$ such that $\sigma_A$ is equal (or as close as numerically possible) to the desired value of $\sigma_A$. This final step 'fixes' the values, and thus allows one to obtain the moments of the abundance distributions as a function of $\sigma_A$ for some fixed $\mu_A$ and $\rho_A$.

Once the solutions have been obtained, recall that the functional form of the abundance distribution is the sum of a (discrete) $\delta$-distribution at 0, and a rectified distribution for strictly positive abundances [Eq. [A4.27]]. That is to say,

$$p(x) = (1 - \phi)\delta(x) + p^+(x), \qquad [A4.32]$$

---

[†]Note that this steady state does not need to be feasible ($x_i > 0$), but it needs to be a global attractor of the deterministic dynamics. This can be guaranteed under some assumptions on the mean and variance of the interaction coefficients [13, 21], yet the methods here work relatively well even when these assumptions are relaxed.

Johannes Nauta, Kaitlin A. Schaal, Ying-Jie Wang, James P.J. Hall, Shai Pilosof, Manlio De Domenico

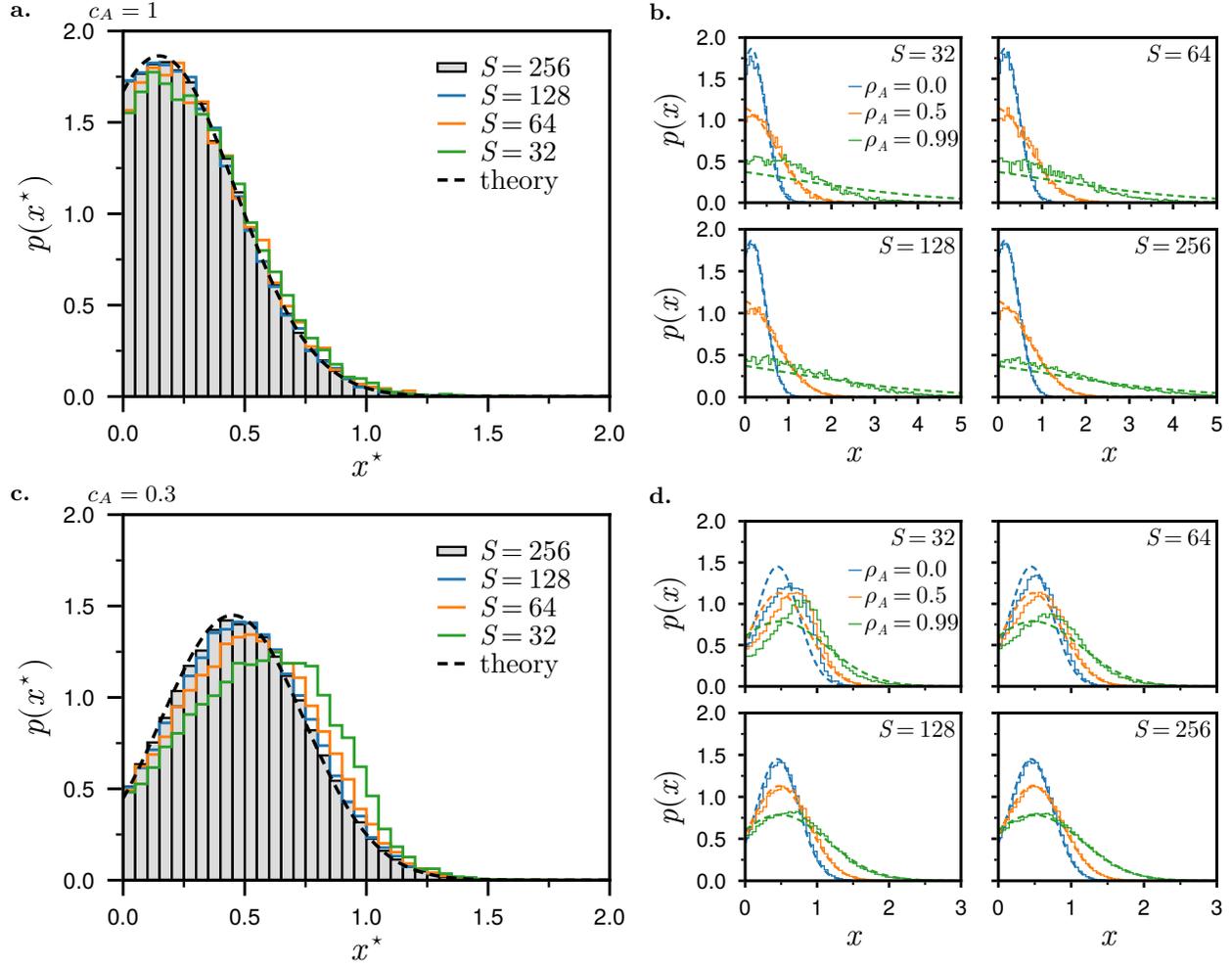

**Fig. S2. Species abundance distributions (SADs) for the generalized Lotka-Volterra system.** We considered systems with $S$ species and random interactions for two interaction densities $c_A = 1$ and $c_A = 0.3$. **a.** Histograms of abundances of non-extinct species with $x_i^* > 0$ for different system sizes and dense interactions with $c_A = 1$. When the system size increases, the theoretical approximation becomes more accurate, yet for relatively small system sizes the main components of the Gaussian distribution are clearly visible. **b.** Same histograms but for different interaction correlation $\rho_A = \text{Corr}(A_{ij} A_{ji})$. As we consider $\mu_A, \rho_A > 0$, increases in $\rho_A$ lead to more competitive interactions. **c,d.** The same, but now for less dense interaction networks with $c_A = 0.3$. Note that mean abundances are typically higher. Other relevant parameters are $\mu_A = 4$, $\sigma_A = 1$, $r_i = 1$, and $C_i = 1$. Histograms are computed across $512$ distinct runs (seeds). Note that the theoretical approximation for $\rho_A$ not too large works surprisingly well, even when system sizes are small, yet that this effect appears to depend on the interaction density $c_A$. For systems with less interactions (lower $c_A$), the approximation is less accurate than for denser ones.

223 where

224 $$\phi = \int_0^\infty p^+(q) \, dq \qquad \text{[A4.33]}$$

225 the fraction of species that survived (i.e., those with positive abundances in the steady state). Note that $\phi$ can also be expressed
226 as a function of $\Delta$, as one readily obtains that $\phi(\Delta) = w_0(\Delta)$, and thus it can be computed essentially for free when finding
227 the parametric solution as mentioned above. The functional form of the rectified Gaussian reads

228 $$p(x) = \frac{1}{\sqrt{2\pi\sigma_0^2}} \exp\left(-\frac{(x - \mu_x)^2}{2\sigma_x^2}\right) \Theta(x) \qquad \text{[A4.34]}$$

229 with $\Theta(x)$ the Heaviside function, and $\mu_x$ and $\sigma_x$ the mean and variance of the unclipped Gaussian, which are defined by

230 $$\mu_x = \frac{1 + \mu_A M_1}{1 - \rho_A \chi \sigma_A^2} \qquad \text{[A4.35a]}$$

231 $$\sigma_x = \frac{\sigma_A \sqrt{M_2}}{1 - \rho_A \chi \sigma_A^2} \qquad \text{[A4.35b]}$$

232 These moments are enough to predict maintenance probabilities of transmissible elements [Eq. (A3.14)], as they define the
233 abundance distributions which are necessary to compute the (moments of) the basic reproduction number.

To illustrate the effectiveness of this approach to approximate abundance distributions for finite systems sizes $S$, we plot the abundances of the persistent species with $x_i^* > 0$ in Fig. S2a. From these results, it becomes clear that abundance distributions are indeed defined by the rectified Gaussian, and that the parametric scheme reproduces the correct moments of this distribution. Notably, even for relatively small system sizes, the solution (which technically holds in the thermodynamic limit for $S \to \infty$) is surprisingly accurate. The mean and variance of the interaction coefficients determine the moments of the distribution over species abundances [Eq. (A4.35)]. Higher mean interaction strengths lead to lower mean abundances, and higher variance lead to more extinctions [13].

**4A.1. Correlations and sparsity in the generalized Lotka-Volterra system.** Let us discuss here in more detail correlations in the interaction matrix. These are defined as

$$\rho_A = \text{Corr}(A_{ij}, A_{ji}) = \frac{\text{Cov}(A_{ij}, A_{ji}) - \mu_A^2}{\sigma_A^2} \tag{A4.36}$$

with $\text{Cov}(A_{ij}, A_{ji}) = \langle A_{ij} A_{ji} \rangle$ the covariance. These correlations typically 'strengthen' the type of interaction, where negative correlations lead to more predator-prey type interactions, and positive ones relate to mutualism or competition. As we consider $\mu_A > 0$ (and $\sigma_A$ sufficiently small), positive correlation here corresponds to increases levels of competition.

Interaction sparsity is defined from a network-perspective, where the interaction network is typically a random network (yet other networks have been considered, such as modular and nested ones) with *connectance* $c_A$. The connectance determines the probability of two species interacting, and hence with probability $1 - c_A$ corresponding components $A_{ij}$ and $A_{ji}$ are 0, and with probability $c_A$ they are sampled from the desired distribution (with correlation $\rho_A$). Generally, species interaction networks are relatively sparse, but one should note that the above derivation of the abundance distributions holds only when the total number of interactions (i.e., the *degree* of a species) scales with $S$. In "truly sparse" interaction networks, where the degree is constant (i.e., does *not* change as $S$ changes), other phenomena arise [23, 26]. We will consider the former case, and the rectified Gaussian distribution is found to correctly approximate abundance distributions for any $c_A > 0$ and different $\rho_A$ (Fig. S2b). Note that for any $c_A \leq 1$ the effective mean and variance change as $\mu_A \to c_A \mu_A$ and $\sigma_A^2 \to c_A \sigma_A^2$, which can be substituted into [Eq. (A4.35)] to obtain the abundance distribution.

**4A.2. The effect of dilution of abundance distributions.** The results described above hold only in the absence of death or dilution. While the effects of its inclusion are subtle, they may result in significant changes both in the abundance distribution and the maintenance of the transmissible elements. For example, species that previously survived the dynamics with low abundance will now be more likely to go extinct. In addition, as the careful reader might have noticed, not including dilution leads to the cost that the transmissible element has on its host to not influence maintenance probabilities at all — while empirical evidence is available that indicates that this cost can be significant [27–30]. The intuitive reason for the fact that it has no effect is that the above reasoning applies to the steady state of the system, which is achieved in the large time regime which is independent of the growth rate. As costs typically reduce growth rate, these effects become negligible if the species are given long times to establish themselves. To illustrate the effect that dilution has on maintenance probabilities, let us consider a very simple simple with $S = N = 1$, and consider the fitness effect $\alpha > 0$ which reduces the growth rate of a specific species hosting the pathogen element. Using Eq. (1), we can easily write down the dynamics for the element-free and element-carrying subpopulations $x$ and $y$, which read

$$\frac{dx}{dt} = \frac{rx}{C}(C - x - y) - dx - \gamma xy + \omega y, \tag{A4.37a}$$

$$\frac{dy}{dt} = \frac{\alpha ry}{C}(C - x - y) - dy + \gamma xy - \omega y, \tag{A4.37b}$$

where for the ecological function we have again the Lotka-Volterra system — which for a single species defines logistic growth — and have introduced a death rate $d$, infection rate $\gamma$, and recovery rate $\omega$.

Using the next-generation matrix formalism, we compute the basic reproduction number as the eigenvalues of $\mathbf{K}$ following the steps in Appendix 2, which lead to the following constraint on the infection rate above which the element becomes endemic [Eq. (A3.25)],

$$\gamma > \gamma_c = \frac{\omega r + (1 - \alpha) rd}{C(r - d)} \tag{A4.38}$$

Hence, for $\gamma > \gamma_c$ we find that $R_0 > 1$, and thus the element with fitness cost $\alpha$ remains endemic. A comparison with a system without dilution illustrates that $\gamma_c$ increases as the cost of carrying the element increases (decreased $\alpha$). This is additionally illustrated in Fig. S3a, where we have plotted the abundance of the infected subpopulation against the fitness effect and the infection ratio $\beta = \gamma / \omega$. When there is no fitness effect on the host ($\alpha = 1$), we find that $\beta_c = \gamma_c / \omega = 1$, yet when the cost of hosting increases (decreases in $\alpha$), the infection ratio $\beta_c \gg 1$, and even for relatively moderate values (e.g., $\alpha = 0.9$) one can see that $\beta_c$ is about an order of magnitude larger than 1. This suggests that transmissible elements that confer a non-zero cost on their hosts must compensate for this by (drastically) increasing their infection rates compared to the rates at which they are lost. This reasoning holds as well in the large and disordered system we considered in the main text as now the average infection ratio $\bar{\beta}$ needs to increase by at least an order of magnitude when the cost of of hosting the plasmid is distributed about $\bar{\alpha} = 0.9$ (compared to no fitness effect for $\bar{\alpha} = 1$, Fig. S3b).

Finally, note that in the simple system the element becomes beneficial for $\alpha > 1$ and, depending on the parameters, maintenance becomes essentially guaranteed for *any* positive infection rate. In fact, when the numerator of Eq. (A3.25) becomes

Johannes Nauta, Kaitlin A. Schaal, Ying-Jie Wang, James P.J. Hall, Shai Pilosof, Manlio De Domenico

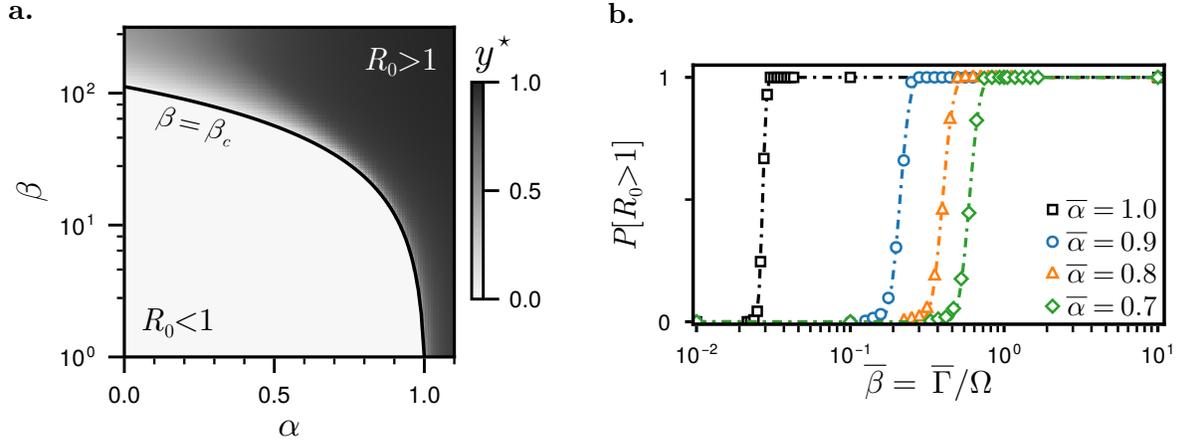

**a.**

$y^\star$

$R_0 > 1$

$\beta = \beta_c$

$R_0 < 1$

**b.**

$P[R_0 > 1]$

$\bar\alpha = 1.0$
$\bar\alpha = 0.9$
$\bar\alpha = 0.8$
$\bar\alpha = 0.7$

$\bar\beta = \bar\Gamma / \Omega$

**Fig. S3. Plasmid maintenance is affected by the fitness effect $\alpha$ and the infection ratio $\beta = \gamma/\omega$. a.** The abundance of the infected subpopulation in the steady state for a simple system $S = N = 1$. Here, $y^\star$, is only non-zero when $R_0 > 1$, which occurs for $\beta > \beta_c$ [or $\gamma > \gamma_c$, see Eq. (A3.25)], as predicted and indicated by the solid line. Relevant parameters are $r = 1$, $d = 0.1$, $C = 1$, and $\omega = 10^{-3}$. **b.** Plasmid maintenance probabilities versus mean infection rate for distinct average fitness effects $\bar\alpha$. Both infection ratios and fitness effects are distributed log-normally (see main text and [Materials and Methods](#) for more details). Dash-dotted lines are fits of the function defined in Eq. (4) (see main text), while markers indicate results from numerical integration. Other relevant parameters are as in Fig. 3, and $\sigma_\beta = 1$.

zero, any non-zero infection rate will lead to maintenance. As $\omega$ is typically small, one can easily see that $\alpha \gtrsim 1$ can often be understood as the point beyond which the transmissible element is maintained. Intuitively this also makes sense, as there is no ecological reason to get rid of beneficial mobile genetic elements by their hosts.

**4B. Element maintenance does not predict element abundances.** Here we would like to briefly mention that plasmid maintenance does not tell us anything about the actual (relative) abundances of plasmids in the system. That is, even when $P[R_0 > 1] = 1$, it can be that a very small fraction of species are actually hosting the plasmid. This can, for example, be seen by plotting the average fraction of infected species $\langle y_i \rangle$, as done in Fig. [S4](#) for different values of the fitness effect $\bar\alpha$ (as in Fig. [S3b](#) above).

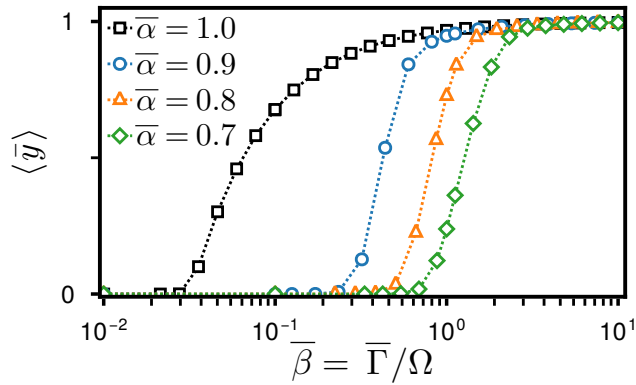

$\bar\alpha = 1.0$
$\bar\alpha = 0.9$
$\bar\alpha = 0.8$
$\bar\alpha = 0.7$

$\langle \bar{y} \rangle$

$\bar\beta = \bar\Gamma / \Omega$

**Fig. S4. Mean plasmid abundance for diluted systems for different fitness effects $\bar\alpha$.** The average fraction of infected species for different mean fitness effects $\bar\alpha$. The fraction is defined as $\langle y \rangle = S^{-1} \sum y_i / (x_i + y_i)$ — hence, for $\langle y \rangle \to 1$, all species consist only of the infected subpopulation (see text). Dotted lines are guides to the eye. Results are shown for systems with $S = 100$ and $d = 0.1$, infection rates and fitness effect are assumed to be log-normal ($\sigma_\beta = 1$, [Materials and Methods](#)), and all other parameters are as in Fig. 3.

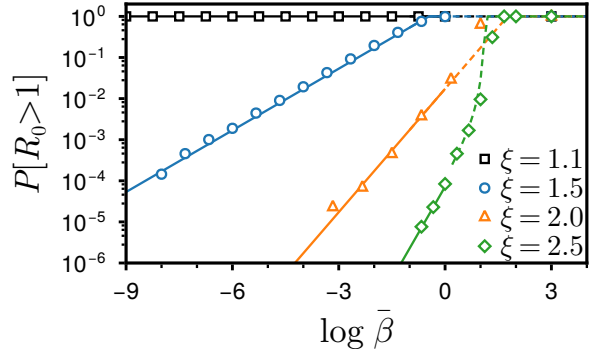

**Fig. S5. Basic reproduction numbers for power-law abundance distributions.** The probability of plasmid maintenance, $P[R_0 > 1]$, in a system with $S = 256$ versus the mean infection ratio $\bar{\beta}$, with $\beta_{ij} \sim \text{LogNormal}(\mu_\beta, \sigma_\beta)$ with mean $\bar{\beta}$ and $\sigma_\beta = 1$ (that is, the parameter $\mu_\beta$ is computed from $\bar{\beta}$ and $\sigma_\beta$). The probability is plotted for several exponents of the power law $p(x) \sim x^{-\xi}$. For $2 < \xi < 3$ the sum $\Sigma_S$ (and thus $R_0$) converges to an $\xi$-stable distribution, for which the maintenance probability can be computed (Eq. A4.41, dashed lines). For $1 < \xi < 2$ the maintenance probability scales as in Eq. (A4.43), yet due to finite size effects the approximation of Eq. (A4.42) tends to work when $\xi \gtrsim 1.2$ (for $\bar{\beta} \ll 1$, solid lines). Note that $P[R_0 > 1] \to 1$ for all $\bar{\beta}$ as $\xi \to 1$, as illustrated by $\xi = 1.1$. For all values of $\xi$, the approximations of Eqs. (A4.42) and (A4.43) remain accurate only when $\bar{\beta} \ll 1$, so above $\bar{\beta} = 1$ we have fitted the $\xi$-stable distribution instead (for $\xi > 2$, green dashed line). Scatter points are results of numerically computing the basic reproduction numbers, while solid and dashed lines are the approximations and $\xi$-stable fits respectively.

**4C. Scaling laws for heterogeneous abundance distributions.** Whereas above we have assumed Lotka-Volterra dynamics, the Gaussian abundance distributions that it gives under the assumed properties do not coincide with empirically observed patterns. Those patterns, in contrast, often predict heterogeneous, or heavy-tailed, distributions for species abundances. This means that $p(x) \sim x^{-\xi}$, for which the normalized distribution for $\xi > 1$ reads[‡]

$$p(x) = \frac{1 - \xi}{\epsilon} \left( \frac{x}{\epsilon} \right)^{-\xi},$$ [A4.39]

which is properly defined for $x > \epsilon$ where $\epsilon > 0$ is some (small) number equal to the minimum value that $x$ can take (31). In such communities the basic reproduction number $R_0$ is no longer well described by a Gaussian sum, and its statistics are controlled instead by the exponent $\xi$.

Let us first consider the homogeneous infection case with a constant infection rate $\beta$, and consider first the case where $2 < \xi < 3$ so that the variance of $p(x)$ diverges, but the mean $\langle x \rangle$ exists and remains finite. Formally, the sum $\Sigma_S = \sum_{i=1}^S x_i$ converges in distribution to a skewed $\xi$-stable law, and the normal (Gaussian) limit is replaced by a $\xi$-stable one. That is to say, we have that

$$\frac{\Sigma_S - \langle x \rangle}{S^{1/(\xi-1)}} \xrightarrow{\text{dist.}} p_\xi(\xi - 1; \theta, \zeta, \eta)$$ [A4.40]

with the arrow denoting convergence in distribution and $P_\xi$ a $\xi$-stable distribution with skewness $\theta$, scale $\zeta$ and location $\eta$. For variates $x_i$ coming from a Pareto distribution [Eq. (A4.39)], we know the mean $\langle x \rangle = \epsilon(\xi - 1)/(\xi - 2)$ and the scale $\epsilon S^{1/(\xi-1)}$, and we know that the distribution is fully right skewed so that the skewness parameter equals unity. Within the context of basic reproduction numbers, which in the homogeneous infection case equals $R_0 = \beta \Sigma_S$, we thus know that their distribution is $p_\xi$. Moreover, we can compute the maintenance probability $P[R_0 > \kappa]$ as

$$P[R_0 > \kappa] = 1 - F_\xi(z) \qquad \text{where} \quad z = \frac{1/\beta - S\epsilon(\xi-1)/(\xi-2)}{\epsilon S^{1/(\xi-1)}}$$ [A4.41]

and $F_\xi(z)$ the cumulative distribution function of the stable distribution $p_\xi$ with skewness, scale and location parameters $\theta = \zeta = 1$ and $\eta = 0$. In this case, the probability can also be approximated as one realizes that for $S \gg 1$ and $\kappa/\beta \gg 1$ the probability is determined by a select few species that very abundant (relative the the remainder of the community) and hence one approximates

$$P[R_0 > \kappa/\beta] \approx SP[x_i > \kappa/\beta] = S \left( \frac{\kappa}{\epsilon \beta} \right)^{1-\xi}$$ [A4.42]

Inspecting this at the infection threshold $\kappa = 1$ gives $P[R_0 > 1] \sim S\beta^{\xi-1}$. This approximation shows that increases in species richness, mediated by $S$, or a modest rise of the infection ratio $\beta$ can outweigh a sub-critical average $R_0$. Ecologically this means that a handful of extremely abundant "super-host" species can ensure plasmid persistence, even when the remaining bulk of the community would not be able to sustain the plasmid on its own.

When the tail of the Pareto distribution is even heavier such that the mean also diverges, which occurs for $1 < \xi < 2$, the sum $\Sigma_S$ is dominated by the single largest term, and extreme-value arguments predict that (32)

$$P[R_0 > \kappa/\beta] \approx 1 - \left[ 1 - \left( \frac{\kappa}{\beta} \right)^{1-\xi} \right]^S$$ [A4.43]

---

[‡]Where for brevity we have let $x_i = x_i^\star$, and assume that $p(x_i) = p(x)$.

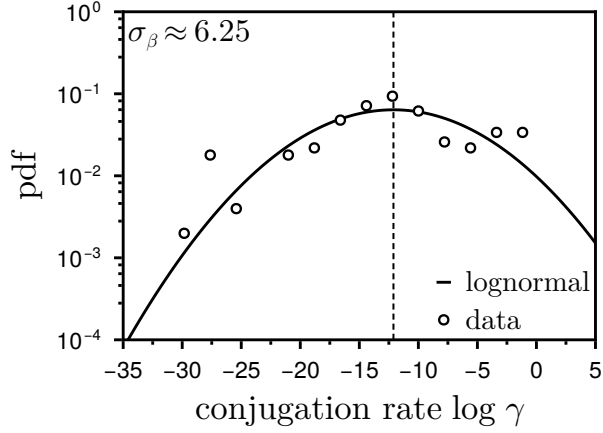

**Fig. S6. Conjugation rates of plasmids are approximately log-normal.** Mean conjugation (infection) rates from (16) are plotted and a log-normal distribution is fitted to the data (see text and Materials and Methods). Vertically dashed line indicates the mean conjugation rate $\bar{\gamma} \approx 10^{-12}$ (which in our work can be considered the infection ratio $\bar{\beta}$), but due to the high variance outliers up to $\gamma \approx 10^{-1}$ are not uncommon (see text).

This maintenance probability rapidly approaches unity as either $S$ or $\beta$ grows, making the average basic reproduction number effectively meaningless. It is worth noting, however, that most often natural abundance distributions do not tend to fall within this regime.

When the infection ratio follows some distribution $p_\beta$ that has finite mean and variance, the above approximations remain valid as a mean field approximation after letting $\beta \to \bar{\beta}$. A comparison between the approximations and numerical computations for log-normally distributed infection ratios is given in Fig. S5.

## 5. On the distributions of plasmid conjugation rates

Our assumption that the infection ratios $\beta_{ij}$ are log-normally distributed is based on data on conjugation rates collected in a diverse set of environments (16). The data of this study contains rates of horizontal gene transfer for a diverse set of MGEs and host species. Akin to our disordered approach, we let the data become species-agnostic by aggregating all species and environments into a single dataset of conjugation rates, and compute the distribution over them. Note that the data contains only information about mean conjugation rates, computed from repeated measurements, and thus these rates and their distribution should be interpreted as an average distribution. Regardless, we plot the distribution of conjugation rates from that study and they appear to be described by a log-normal distribution relatively well (Fig. S6). In addition, as the variance for the found value of $\sigma_\beta$ can be approximated as $\text{Var}(\beta_{ij}) \propto e^{2\sigma_\beta^2}$, it suggests that the infection rates tend to spread widely and occasional large outliers are increasingly common. Note that this aligns with our suggestions illustrated in Fig. 5, in that large variances in infection rates may underlie plasmid maintenance. Whereas a full statistical examination of the distribution is considered to be out of the scope of the work presented here, we believe that this simple analysis justifies the use of the log-normal distributions for infection rates and illustrates the effect that random samples may induce plasmid maintenance.

## 6. Possible extension to multiple mobile genetic elements

Finally, we sketch the outline of how one would extend our results to cases with multiple mobile genetic elements ($N > 1$). To do so, let us consider systems the limiting case where no transmissible element can be co-hosted with other elements — that is, host species cannot become co-infected. Other cases are unlikely to be handled well by our generic model, at least analytically, and other models, such as those put forward in (33–35). In this limiting case, the formulation of Eq. (A1.4) simplifies greatly, as one obtains instead

$$\frac{dx_i^\emptyset}{dt} = x_i^\emptyset f_i(\boldsymbol{x}) - x_i^\emptyset \sum_{j,u} \widetilde{\Gamma}_{ij}^u x_j^u + \sum_u \Omega_i^u x_i^u \qquad \text{[A6.44a]}$$

$$\frac{dx_i^u}{dt} = x_i^u f_i(\boldsymbol{x}) + x_i^\emptyset \sum_j \widetilde{\Gamma}_{ij}^u x_j^u - \Omega_i^u x_i^u \qquad \text{[A6.44b]}$$

Again, note that for $f_i(\boldsymbol{x}) = 0$ one finds that

$$\frac{dx_i^\emptyset}{dt} + \sum_u \frac{dx_i^u}{dt} = 0 \qquad \text{[A6.45]}$$

so that infection- and background-processes do not change abundances, as required. Let us focus on the infected subsystems by defining $\boldsymbol{y}^u = (x_1^u, \ldots, x_S^u)$ and then defining $\boldsymbol{y} = (\boldsymbol{y}^1, \ldots, \boldsymbol{y}^{S \cdot 2^N})$. Under these conditions, we can again write the vectors $\boldsymbol{g}$

and $\boldsymbol{h}$ as

$$g_i^u = x_i^\emptyset \sum_j \Gamma_{ij}^u y_j^u, \tag{A6.46a}$$

$$h_i^u = \Omega_i^u x_i^u - x_i^u f_i(\boldsymbol{x}) \tag{A6.46b}$$

and the matrices $\mathbf{G}$ and $\mathbf{H}$ with elements

$$G_{ij}^u \equiv \frac{\partial g_i^u}{\partial x_j^u} = \widetilde{\Gamma}_{ij}^u x_i^\emptyset \tag{A6.47a}$$

$$H_{ij}^u \equiv \frac{\partial h_i^u}{\partial x_j^u} = \left[\Omega_i^u - f_i(\boldsymbol{x}) - x_i^u \frac{\partial f_i(\boldsymbol{x})}{\partial x_j^u}\right]_{\boldsymbol{y}=0} = \Omega_i^u \tag{A6.47b}$$

where again the last equality is because $f_i(\boldsymbol{x}) = 0$ when $\boldsymbol{y} = 0$, by definition. These matrices are now block-matrices;

$$\mathbf{G} = \begin{pmatrix} \mathbf{G}^1 & & \\ & \ddots & \\ & & \mathbf{G}^N \end{pmatrix}, \qquad \mathbf{H} = \begin{pmatrix} \mathbf{H}^1 & & \\ & \ddots & \\ & & \mathbf{H}^N \end{pmatrix} \tag{A6.48}$$

with elements as in Eq. (A6.47). Note that this means that all $\mathbf{H}^u$ are diagonal matrices themselves, which again allows one to easily obtain $\mathbf{H}^{-1}$, that defines the next-generation matrix

$$\mathbf{K} = \mathbf{G}\mathbf{H}^{-1} \tag{A6.49}$$

for which the eigenvalues, and thus the basic reproduction numbers, can now be readily determined.

However, in contrast with the derivation that lead to expressions for $R_0$ [Eqs. (A3.12) and (A3.24)], we now obtain a set of basic reproduction numbers for each distinct element $R_0^u$. Note that herein lies a shortcoming of the next-generation matrix formalism, as only the largest basic reproduction number is predictive of a pathogen becoming endemic. That is to say, only when

$$\mathcal{R}_0 = \max_u R_0^u > 1 \tag{A6.50}$$

we can reason about the maintenance of element $u$ for which $R_0^u$ is largest. Yet, if there are other elements with an $R_0^v > 1$, yet $R_0^v < R_0^u$, we will need to check whether element $v$ can invade a system wherein $u$ is endemic. As more basic reproduction numbers are larger than zero, one should verify all different combinations of invasions, which quickly becomes intractable. This issue illustrates that, while our model is generic, tackling it using the next-generation matrix is limited in scope and does not readily apply to systems with large numbers of transmissible elements or pathogens.

%

\end{document}